\documentclass[%
reprint,
nofootinbib,
 amsmath,amssymb,
 aps,
prd,floatfix,
nobalancelastpage%
]{revtex4-1}
\usepackage{isotope}
\usepackage{bbm}
\def\Ars{$\isotope[40]{A\lowercase{r}}$}
\def\Arl{${\isotope[40][18]{A\lowercase{r}}}$}
\def\Ar40{\Ars}
\usepackage{epsfig}

\def\Journal#1#2#3#4{{#1} {#2} (#4) #3 }
\def\EPJA{{\em Eur. Phys. J.} A}

\def\NPA{{\em Nucl. Phys.} A}

\def\PLB{{\em Phys. Lett.} B}

\def\PREP{\em Phys. Rep.}

\def\PRC{{\em Phys. Rev.} C}

\def\ss{\mbox{\boldmath $\sigma$}}
\newcommand{\be}{\begin{equation}}
\newcommand{\ee}{\end{equation}}
\newcommand{\bea}{\begin{eqnarray}}
\newcommand{\eea}{\end{eqnarray}}

\DeclareMathOperator{\IM}{Im}

\usepackage{graphicx}
\usepackage{dcolumn}
\usepackage{bm}

\begin{document}

\title{Progress and Simulations for Intranuclear Neutron--Antineutron\\ Transformations in \boldmath\Arl\unboldmath}
%
\author{Joshua L. Barrow}
 \email{jbarrow3@vols.utk.edu}
 \affiliation{The University of Tennessee at Knoxville, Department of Physics and Astronomy, 1408 Circle Drive, Knoxville, TN 37996, USA}
 \altaffiliation[Also at ]{Fermi National Accelerator Laboratory}
 
\author{Elena S. Golubeva}
 \email{golubeva@inr.ru}
 \affiliation{Institute for Nuclear Research, Russian Academy of Sciences, Prospekt 60-letiya Oktyabrya 7a, Moscow, 117312, Russia}
 
\author{Eduard Paryev}
 \email{paryev@inr.ru}
 \affiliation{Institute for Nuclear Research, Russian Academy of Sciences, Prospekt 60-letiya Oktyabrya 7a, Moscow, 117312, Russia}
 
\author{Jean-Marc Richard}%
 \email{j-m.richard@ipnl.in2p3.fr}
 \affiliation{Institut de Physique des 2 Infinis de Lyon, Universit\'e de Lyon, CNRS-IN2P3--UCBL, 4 rue Enrico Fermi, Villeurbanne, France}




\date{\today}
\begin{abstract}
With the imminent construction of the Deep Underground Neutrino Experiment (DUNE) and Hyper-Kamiokande, nucleon decay searches as a means to constrain beyond Standard Model (BSM) extensions
are once again at the forefront of fundamental physics. Abundant neutrons within these large experimental volumes, along with future high-intensity neutron beams such as the European Spallation Source, offer a powerful, high-precision portal onto this physics through searches for $\mathcal{B}$ and $\mathcal{B}-\mathcal{L}$ violating processes such as neutron--antineutron transformations ($n\rightarrow\bar{n}$), a key prediction of compelling theories of 
baryogenesis. With this in mind, this paper discusses a novel and self-consistent intranuclear simulation of this process within \Arl, which plays the role of both detector and target within DUNE's gigantic liquid argon time projection chambers. An accurate and independent simulation of the resulting intranuclear annihilation respecting important physical correlations and cascade dynamics for this large nucleus is necessary to understand the viability of such rare searches when contrasted against background sources such as atmospheric neutrinos. Recent theoretical improvements to our model, such as the first calculations of the \Arl~ intranuclear radial annihilation probability distribution and the inclusion of a realistic $\bar{n}A$ potential, are discussed. A Monte Carlo simulation comparison to another publicly available $n\rightarrow\bar{n}$ generator within GENIE is shown in some detail. The first calculation of \Arl's $n\rightarrow\bar{n}$-intranuclear suppression factor, an important quantity for future searches at DUNE, is also completed, finding $T_R^{Ar}\sim 5.6\times 10^{22}\,\mathrm{s}^{-1}$.
\end{abstract}

\maketitle


\section{\label{sec:intro}Introduction}
\subsection{\label{sec:TheoryBackground}Theoretical Background}
Baryon number ($\mathcal{B}$) violation is the only remaining component of the Sakharov conditions \citep{Sakharov:1967dj} which has yet to be confirmed experimentally, an innate requirement for proper baryogenesis. While it was shown \citep{tHooft:1976rip} that $\mathcal{B}$ itself is not an exact symmetry of the Standard Model (SM) and is only \textit{infinitesimally} violated via $\mathcal{B}-\mathcal{L}$-conserving non-pertubative electroweak instanton processes, this marginal infraction does not appear to naturally explain the observed excess of matter over antimatter. Thus, it is possible that not only must $\mathcal{B}$ be violated, but that $\mathcal{B}-\mathcal{L}$ must too be broken via some beyond Standard Model (BSM) mechanism.

Such explicit $\mathcal{B}$ violating extensions can be constructed as particular types of (di)nucleon decays, mainly within the structures of dimension $d=6$ and $d=9$ operators. While it is believed that many $d=6$ operators (governing things such as proton decay) may be heavily (or altogether) suppressed \citep{Nussinov:2001rb,Arnold:2012sd}, this may not be the case for $d=9$ operators with the basic structure $c \cdot \,qqq\bar{q}\bar{q}\bar{q}/M^5$, where the high-mass scale $M$ can be rather low at perhaps $\sim100\,$TeV. These operators allow for modes of dinucleon processes of disappearance or \textit{oscillation}, a most important member of which is neutron--antineutron transformation ($n\rightarrow\bar{n}$). Popular (modern and minimal) BSM extensions \citep{Kuzmin:1985mm,Babu:2006xc,Babu:2013yca,Grojean:2018fus} permitting such a process can dynamically create a proper baryon abundance in the early universe, even while predicting a reasonable and possibly observable \textit{upper} limit on the mean transformation time in vacuum, allowing the theory to be well constrained if not eventually entirely eliminated experimentally. This prospect seems stronger than ever given new studies from lattice quantum chromodynamics calculations \citep{Rinaldi:2018osy} and other interesting, highly general recent works \citep{Anca,Grojean:2018fus,Nesvizhevsky:2018tft,Berezhiani:2018pcp}.

\subsection{\label{sec:DUNEintrannbar}DUNE and Intranuclear \boldmath$n\rightarrow\bar{n}$ in \Arl\unboldmath}
The Deep Underground Neutrino Experiment (DUNE) will be the future heart of American particle physics. It will contain $40\,$kt of liquid argon (\Arl) inside its
fiducial mass, acting both as prospective neutrino target and ionization detection medium within a set of four gargantuan time projection chambers. While its main goals of study are eponymous, some nuclear and particle physicists see DUNE as the next large-scale step toward answering arguably more fundamental and controversial questions about the nature of matter and the laws which govern it and its origins. Of course, BSM physics searches like $n\rightarrow\bar{n}$ fit within this program well.

This oscillatory or transformational process can be best encapsulated within a single value: the mean vacuum (free) transformation period, $\tau_{n\rightarrow\bar{n}}$. Taking account of the bound nature of the intranuclear $n$, the inclusion of a so called intranuclear suppression factor $T_R$ converts this free transformation time into a bound transformation time $T=T_R\,\tau_{n\rightarrow\bar{n}}^2~$, where $T_R$ is quite large. Some details of this derivation, along with the first calculation of this factor for the \Arl{}  nucleus, will be described in some detail in later sections.

Previous searches for $n\rightarrow\bar{n}$ have been carried out using both free \citep{BaldoCeolin:1994jz} and bound \citep{Jones84,Takita86,Berger90,Chung02,Abe15,Bergevin10,Aharmim:2017jna} $n$'s. The most successful thus far has been Super-Kamiokande's \citep{Abe15} bound water-Cherenkov $n\rightarrow\bar{n}$ search within \isotope[16][8]{O}, which, given $24$ candidate events and an expected background of $24.1$ atmospheric neutrino events over $\sim 4$ years within $22.5$ kT of fiducial mass, set a limit on the intranuclear $n$ lifetime of $1.9\times10^{32}$ years, which, when converted to a free transformation time, became $\geq 2.7 \times 10^8s$. The best free $n$ search occurred at the Institut Laue-Langevin \citep{BaldoCeolin:1994jz}, and puts a lower limit at roughly the same order of magnitude with no apparent backgrounds.

\boldmath\subsection{\label{sec:PastMCWork}Past Intranuclear $n\rightarrow\bar{n}$ Simulation Work}\unboldmath

It is critical to recognize the interdependency of the computational modeling of BSM signals \textit{and} backgrounds in the estimation of detector efficiencies and background rates within the context of large modern experiments. Thus, it becomes crucially important that one should take care to model BSM signals and backgrounds as completely, consistently, and rigorously as necessary by employing limited approximations which attempt to preserve as much physics as possible. This is especially true given nontrivial automated triggering schemes planned for future rare event searches. Unfortunately, in the case of many previous $n\rightarrow\bar{n}$ studies, this has not altogether been the case.

While previous work in \Arl ~by Hewes \citep{Hewes:2017xtr} and others from the DUNE Nucleon Decay Working Group using newly-constructed modules within the GENIE \citep{Andreopoulos:2015wxa} Monte Carlo (MC) event generator has made excellent progress and developed technically fantastic analysis schemes using convolutional neural networks and boosted decision trees, many of the underlying physics assumptions of a hypothesized $n\rightarrow\bar{n}$ signal within the GENIE \textit{default} model are not entirely correct. Some approximate notions include:

\begin{enumerate}
    \item The assumption that the annihilation occurs along the nuclear density distribution of the nucleus, as discussed in \citep{Hewes:2017xtr} and other works, even while the outcomes of \citep{Friedman:2008es} are referenced openly and often used; this ignores the surface dominance of the transformation and subsequent annihilation, a key prediction of \citep{Dover:1996ee,Friedman:2008es}. Using two models (including a more holistic quantum-mechanical one similar in spirit to \citep{Friedman:2008es}), these two assumptions are tested, showing moderate disagreement.
    \item Employing a single nucleon momentum distribution described by a \textit{nonlocal}, relativistic Bodek-Ritchie \citep{Bodek:1980ar,Andreopoulos:2015wxa} Fermi gas (the radial dependence of the annihilating (anti)nucleons' momenta are ignored).
    \item Only $\sim 10$ annihilation channels (a la \citep{Abe15}) are assumed to be necessary to describe the annihilation products. This seems low, as $\sim 100$ are known, many of them containing heavier resonances; these heavier species can be responsible for $\sim 40 \%$ of all pion ($\pi^{0,\pm}$) production.
    \item A true cascade model has not yet been employed, and instead has been approximated as a single effective interaction (GENIE's Intranuke hA2015 \citep{Andreopoulos:2015wxa} was used for previous results)
    \item There also exists no de-excitation model(s) (nucleon evaporation, etc.) within current publicly available builds of GENIE for \Arl
    \item No comparison tests against antinucleon annihilation data have yet been considered.
    \item Only an rough estimation of the nuclear suppression factor of \Arl ~has thus far been used to approximate lower limits on the transformation period.
\end{enumerate}

It is no doubt that some of these current technicalities proceed directly from the secondary nature of the GENIE $n\rightarrow\bar{n}$ module's genesis, a consequence of GENIE's top-down structure and first-and-foremost focus on neutrino interactions. This being said, similar issues or inconsistencies are known to exist in other work \citep{Gustafson16}, and detailed explanations of past simulation's internal processes are quite lacking \citep{Gibin:1996ig,Abe15}, but to individually contend these here is not the goal of this article.

\boldmath\subsection{\label{sec:ThisMCWork}This Work}\unboldmath
Instead, a new and antinucleon-data-tested model is offered here which does not ignore or grossly approximate these effects. In this work, first, the intranuclear oscillation and subsequent annihilation of an $\bar{n}$ is considered using the Sternheimer equation and a proper handling of neutron wave functions. For each neutron shell of \Arl, we calculate the induced $\bar{n}$-component by solving the Sternheimer equation with a realistic $\bar{n}A$ interaction, as done in previous works by Dover, Gal, and Richard. From this, the shell-by-shell and average radial annihilation probability distributions are derived, the first calculation of its kind for \Arl~and now an integral part of the Monte Carlo simulations discussed later in this work. As a consequence of this study, the first calculation of the intranuclear $n\rightarrow\bar{n}$ transformation suppression factor for \Arl~has been completed, finding a value of $T_R^{Ar}\sim 5.6\times 10^{22}\,\mathrm{s}^{-1}$, some $\sim12\%$ \textit{smaller} than previous coarse estimates taken from \isotope[56]{Fe} \citep{Hewes:2017xtr}. Secondly, the first discussion and implementation of novel annihilation pair dynamics in the presence of a realistic $\bar{n}$ potential are completed, where mass defects and momentum modification have now been taken into account within these simulations in a self-consistent way. Third, we consider how changes to our model compare to experiment and past simulation work in extranuclear $\bar{p}$\isotope[12]{C} annihilation at rest \citep{Golubeva:2018mrz} before moving on to consider extranuclear $\bar{n}C$, and then intranuclear $n\rightarrow\bar{n}$ and $\bar{n}N$ annihilation within \Arl. Finally, we analyze our investigation and slight perturbation of GENIE's publicly available $n\rightarrow\bar{n}$ module, attempting to understand the commonalities and differences across both simulations; critically, a discussion is begun as to how ignorance of some particularly physically relevant correlations may affect the eventual feasibility of a true event's observation in DUNE.

\boldmath\section{\label{sec:intranuclear}The Intranuclear $\bar{n}$ Lifetime of Deuterium and \Ar40}\unboldmath

\subsection{\label{sec:concepts}Concepts and Pertinent Questions}

Assuming that the $n\rightarrow \bar{n}$ transformation does occur in a vacuum, an interesting question arises: what are the consequences for a nucleus? Of course, if an intranuclear $n$ becomes an $\bar{n}$,
an annihilation will eventually take place within the nucleus, releasing $\sim 2$\,GeV of rest-mass energy, from which $\sim 4\mbox{-}5$ mesons are emitted on average. After this point, the wounded nucleus will evaporate several nucleons and perhaps break into unstable daughter nuclei. Several questions are immediately raised:
\begin{enumerate}
 \item When a $n$ tentatively becomes an $\bar{n}$, it ceases feeling a smooth potential of $\lesssim 50$\,MeV and instead experiences a (complex) potential whose magnitude is $\gtrsim 100$\,MeV. How much is such a transformation \textit{suppressed} by this change in potential?
 \item A deep annihilation could produce multiple fragments with the primary mesons ultimately being absorbed. Alternatively, a peripheral annihilation \citep{Dover:1996ee} would probably release a large fraction of the primary mesons and at most rip out only a few nucleons, albeit with a more asymmetric topology. So, where, \textit{preferentially}, does the annihilation take place?
 \item Many measurements were accumulated at Brookhaven throughout the 1960-70s, and later at CERN thanks to the LEAR facility (1982-1996), which benefited from a pure, intense, and cooled antiproton ($\bar{p}$) beam. For a review, see, e.g., \citep{Amsler:1991cf,Klempt:2002ap}. However, LEAR was shared by many experiments with various aims dealing with fundamental symmetries, strangeness physics, exotic mesons, etc., and experiments providing knowledge of vital systematics for $\bar{N}N$ and $\bar{N}A$ measurements were not given top priority. With this in mind, \textit{is} our knowledge of antinucleon-nucleon ($\bar{N}N$) and antinucleon-nucleus $(\bar{N}A)$ interactions \textit{sufficient} to carry out such an investigation?
 \item Some concerns have been expressed about the reliability of the estimate of the $n\rightarrow\bar{n}$ lifetime inside nuclei within a straightforward nuclear-physics approach; see, e.g., \citep{Costa:1983wc,Oosterhof:2019dlo}. Can one face these criticisms and demonstrate stable and consistent results?
\end{enumerate}
In this section, a review of these questions is discussed with methods based on the Sternheimer equation (see, e.g., \citep{1970PhRvA...1..321S}, and refs.\ therein), as used by Dover et al.\and Friedman et al.\  \citep{Dover:1982wv,Dover:1985hk,Dover:1994ha,Friedman:2008es} throughout past discussions of these topics. Such methods are then applied to the \Ar40 isotope, which comprises the main component of the DUNE detector. For completeness, the main steps of these calculations are repeated.

\subsection{\label{sec:deuteronlifetime}Lifetime of the Deuteron}

As a \textit{brief warm-up}, consider a simplified deuteron, consisting of a pure $s$-wave bound state of a proton ($p$) and neutron ($n$). One may adopt the wave function by Hulthen, which has been tuned to reproduce the correct binding energy and spatial extension. It reads (see, e.g., \citep{Hulthen:220876,Adler:1975ga})
\begin{equation}
\begin{gathered}
 \Psi_n=\frac{1}{\sqrt{4\,\pi}}\,\frac{u_n(r)}{r}~,\quad  \int_0^\infty u_n(r)^2 \mathrm{d}r=1~,\\
 u_n(r)=N_n \,\left[\exp(-\lambda_1\,r)-\exp(-\lambda_2\,r)\right]~,
 \end{gathered}
 \end{equation}
where $N_n$ is a normalizing factor, $r$ is in GeV$^{-1}$, $\lambda_1=0.2316\,\hbar c$ and $\lambda_2=5.98\,\lambda_1$. 
It has been checked that using another wave function does not change the following results significantly, provided it fits the deuteron energy and radius. 

In the presence of $n\rightarrow\bar{n}$ transformations, the wave function becomes
 \begin{equation}
 \frac{1}{\sqrt{4\,\pi}}\,\left[\frac{u_n(r)}{r}\,|pn\rangle+ \frac{w(r)}{r}\,|p\bar n\rangle\right]~.
 \end{equation}

Assuming an arbitrary strength $\gamma=1/\tau$ for the $n\rightarrow \bar n$ transition, the induced $\bar n$ component $w$ is given by the Sternheimer equation
\begin{equation}\label{eq:Stern-deut}
-\frac{w''(r)}{m}+ V\,w(r)-E\,w(r)=-\gamma\,u_n(r)~.
\end{equation}
This gives an exact estimate of the first-order correction to the wave function and, hence, of the second order correction to the energy without involving a summation over the unperturbed states. Here, $E$ is the unperturbed energy and $V$ the antineutron-proton optical potential, resulting in a width
\begin{align}
  \Gamma=&-2\int_0^\infty |w(r)|^2\, \IM V(r)\,\mathrm{d}r\label{eq:width1}\\
 {}=& -2\gamma \int_0^\infty u_n(r)\, \IM w(r)\,\mathrm{d}r\label{eq:width2}~.
\end{align}

This immediately implies the scaling law $\Gamma\propto \gamma^2$, or for the lifetime $T=\Gamma^{-1}$ of the deuteron
\begin{equation} 
 T=T_R\,\tau^2~,
\end{equation}
where $T_R$, sometimes called the ``intranuclear suppression factor'', is actually a \emph{reduced lifetime}. For the optical potential $V(r)$, there are some models that are tuned to reproduce the main features of low-energy antinucleon-nucleon scattering and have predicted the shift and broadening of the low-lying levels of the protonium atom. 
A method to solve Eq.~\eqref{eq:Stern-deut} is discussed in \citep{Friedman:2008es}, which is similar to the one used in the earlier studies \citep{Dover:1982wv,Dover:1985hk}, and involves the matching of several independent solutions corresponding to various limiting conditions. The alternative below directly provides the desired solution. First, to get the neutron wave function from the neutron potential $V_n(r)$, one should solve the radial equation
\begin{equation}\label{eq:neutron-eq1}
-{u_n}''/\mu +V_n(r)\,u_n(r)=E\,u_n(r)~,
\end{equation}
subject to $u_n(0)=u_n(\infty)=0$. A method adapted from aircraft engineering \citep{Multhopp} consists of the change of variables $r=r_0\,x/(1-x)$, where $r_0$ is a typical distance, and, for the wave function $u_n(r)=\tilde u_n(x)$, of  an expansion
\begin{equation}\label{eq:neutron-eq2}
\tilde u_n(x)=\sum_{j=1}^N a_j\, \sin(j\,\pi\,x)~
\end{equation}
in which the coefficients $a_j$ are closely related to the values of the function at the points $x_i=i\,/(N+1)$, for $i=1,2,\dots N$. This results in a $N\times N$ eigenvalue equation $ A\,U_n=E\,U_n$, where $A$ is the discretized Hamiltonian and $U_n$ is the vector of the $(1-x_i)\,\tilde u_n(x_i)$. See, for instance, \citep{Richard:1992uk}, for an application to quarkonium in potential models. For the Sternheimer equation \eqref{eq:Stern-deut}, one gets a simple matrix equation
\begin{equation}\label{eq:neutron-eq3}
(\bar A-E\, \mathbbm{1})\,W=\gamma \,U_n~,
\end{equation}
where the $N\times N$ matrix $\bar A$ is the discretized Hamiltonian for the $\bar{n}$, $E$ the $n$ energy, and $W=\{(1-x_i)\,\tilde w(x_i)\}$ is the vector containing the $\bar{n}$ wave function. This calculation is fast and robust. 

Besides the deuteron energy $E_0=-0.0022\,$GeV and wave function $u_n$, solving Eq.~\eqref{eq:Stern-deut} requires a model for the antineutron-proton potential $V$. As shown by Fermi and Yang \citep{Fermi:1959sa}, the long-range part of the $\bar N N$ potential in isospin $I$ is deduced from the $NN$ interaction in isospin $I$ by the $G$-parity rule: if a meson (or set of mesons) with $G=+1$ is exchanged, it gives the same contribution, while under $G=-1$ exchange, the sign is flipped. The complex short-range part of $V$ is fitted as to reproduce the low-energy data on $\bar{p}$ scattering and protonium. The so-called DR2 model \citep{Dover:1980pd,Richard:1982zr} has been adopted, and it was checked that variants such as the Kohno-Weise potential~\citep{Kohno:1986fk} produce very similar results. Of course, the isospin $I=1$ of the potential is used for the full calculation of the deuteron lifetime.

\begin{figure}[ht!]
 \centerline{%
\includegraphics[width=1.0\columnwidth]{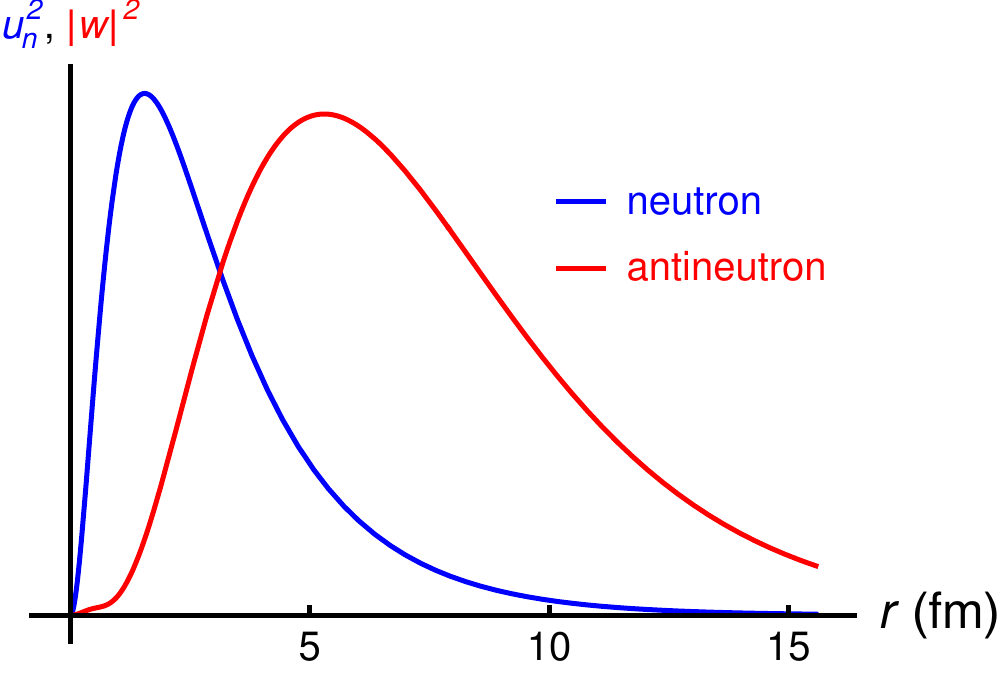}
}
 \caption{The $\bar{n}$ radial distribution (red), arbitrarily rescaled to fit the figure, as compared to the nominal neutron distribution (blue) for the deuteron.
}
 \label{fig:deut}
\end{figure}

However, an annihilation requires both the presence of a neighbor nucleon and an $\bar{n}$, so the annihilation must take place at a slightly less peripheral site than the $\bar{n}$ density itself.  For the deuteron, the reduced lifetime is estimated to be about $T_R^D\simeq 3\times 10^{22}\,\mathrm{s}^{-1}$. This result shows great consistency with other recent calculations such as~\citep{Kopeliovich:2011aa}. A calculation by Oosterhof et al.~\citep{Oosterhof:2019dlo}, based of chiral effective field theory, is in some disagreement, but it has been revisited very recently by Haidenbauer et al.~\citep{Haidenbauer:2019fyd} who found a perfect agreement with the old estimate by Dover et al.~\citep{Dover:1982wv}. 


This estimate is \textit{remarkably stable}. For instance, increasing the core of the $\bar n p$ interaction \textit{by a factor} 10 results in only a $20$\% increase of $T_R$. Even with a large $|V|$, the $\bar{n}$ transformation is more suppressed, but it actually annihilates more efficiently. Remarkably, there is an almost exact cancellation between these two effects. It can also be seen that the calculation is sensitive mainly to the value of $V(r)$ near $0.8-1$\,fm. This is fortunate, as low-energy $\bar{p}$ scattering on nucleons and the shift of the antiprotonic hydrogen atom probes essentially this region and so one cannot determine the interaction at closer distances.\footnote{JMR thanks Femke Oosterhof for discussions on this point.}

In Ref.~\cite{Dover:1982wv}, it is shown that adding a realistic $D$-wave component, and a Sternheimer equation attached to it, does not modify the result significantly.

\boldmath\subsection{\label{sec:argonlifetime}Lifetime of the \Ar40 nucleus}\unboldmath

In Refs.~\citep{Dover:1982wv,Friedman:2008es}, there are estimates of the lifetime of nuclei that were important to analyze some  past underground experiments, in particular  \isotope[16]{O} \citep{Abe15} and  \isotope[56]{Fe} \citep{Chung02}. In the present paper, we study the case of \isotope[40]{Ar} which is relevant for the DUNE experiment. 

The detailed properties of atomic nuclei are well accounted for by sophisticated Hartree-Fock calculations. For many applications, it turns out to be rather convenient to use \textsl{ad-hoc} shell-model wave functions that are tuned to reproduce the main properties of the nuclei, in particular the spatial distribution of $p$'s and $n$'s. This was done in connection with the compilation of nuclear data, see, e.g., \citep{Ajzenberg-Selove:1977gcr}. In the present study, besides a more efficient handling of the Sternheimer equation, our variant consists in using the strategy outlined within and neutron wave functions from \citep{Bolsterli:1972zz}; these wave functions correspond to a fit of the main static properties of the nucleus of interest. The $p$ and $n$ wave functions have been calculated for us by Karim Bennaceur, in the so-called ``filling approximation": the nucleus is supposed to be spherical, implying that the states of each shell are populated with the same (integer or fractional) occupation number. 

The second ingredient of our work is the $\bar n$-nucleus potential. With the noticeable exception of the OBELIX collaboration having studied antineutron scattering \cite{Ableev:1994ac,Iazzi:2000rk}, most data deals with the $\bar p$-nucleus interaction, either via $\bar p$ scattering or antiprotonic atom formation. The question is whether one can reasonably assume that the $\bar n$- and $\bar p$-nucleus potentials are nearly equal. 

The most striking feature of $\bar NN$ cross-sections is the smallness of the charge-exchange component, already stressed by the team having discovered the antiproton \citep{Chamberlain:1957}, and confirmed in further measurements \citep{Klempt:2002ap}.
Indeed, a one-pion-exchange alone would make the charge-exchange  the largest contribution to the total cross section. This implies a large cancellation of the isospin $I=0$ and $I=1$ amplitudes, resulting in a somewhat  isospin-independent $\bar NN$ interaction, and is confirmed by the available experimental data. If one compares the $\bar p p$ and $\bar n p$ total cross sections, they almost coincide, and differ only in tentative extrapolations towards lower energies; see \cite{Iazzi:2000rk,Bugg:1987nq}. This is confirmed by a comparison of the angular distribution of $\bar p$ scattering on the isostopes \isotope[16]{O} and \isotope[18]{O} at the same energy by the PS184 collaboration, with practically no differences \cite{Bruge:1986fd}.

Considering this, we shall soon see in the text which follows that changes in the $\bar n$-nucleus interaction, such as its departure from understood $\bar p$-nucleus interactions, results in only very small modifications of the estimated lifetimes. Of course, the $\bar p$-nucleus potential fitting scattering experiments and antiprotonic-atom data is the strong-interaction part, and Coulomb effects have been adequately removed. 

\begin{figure}[ht!]
 \centerline{\hspace*{-.5cm}
\includegraphics[width=1.0\columnwidth]{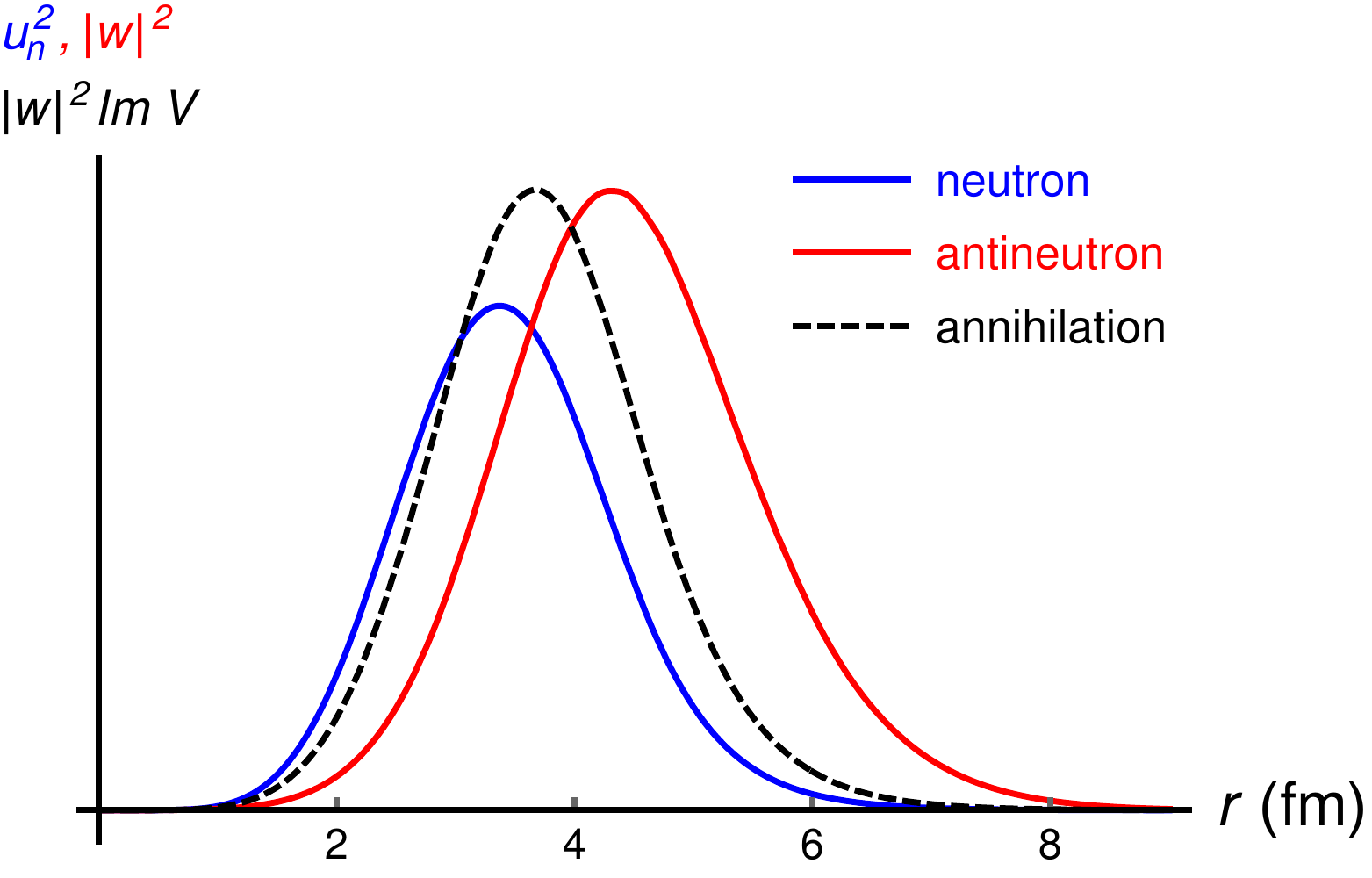}}
 \caption{The $\bar{n}$ radial distribution (red), arbitrarily rescaled to fit the figure, as compared to the nominal neutron distribution (blue), and the annihilation density of \eqref{eq:width2}, also arbitrarily rescaled (dashed black), for the $1d_{5/2}$ shell of \Arl.}
\label{fig:Ar1d52}
\end{figure}

For each $n$ shell, there is an induced $\bar{n}$ wave function governed by an equation analogous to \eqref{eq:Stern-deut}, with a centrifugal term for $p$, $d$ and $f$ states, where $V$ is now the $\bar{n}$-\isotope[39]{Ar} potential, and $m$ the corresponding reduced mass. In Fig.~\ref{fig:Ar1d52}, some details are given for one of the external shells which contributes most to the instability, namely $1d_{5/2}$. It is seen that $\bar{n}$ are produced in the tail of the $n$ distribution, with subsequent annihilation at the surface of the matter distribution; the pattern is similar for all other shells.

\begin{figure}[ht!]
 \centerline{\hspace*{-.5cm}
\includegraphics[width=1.0\columnwidth]{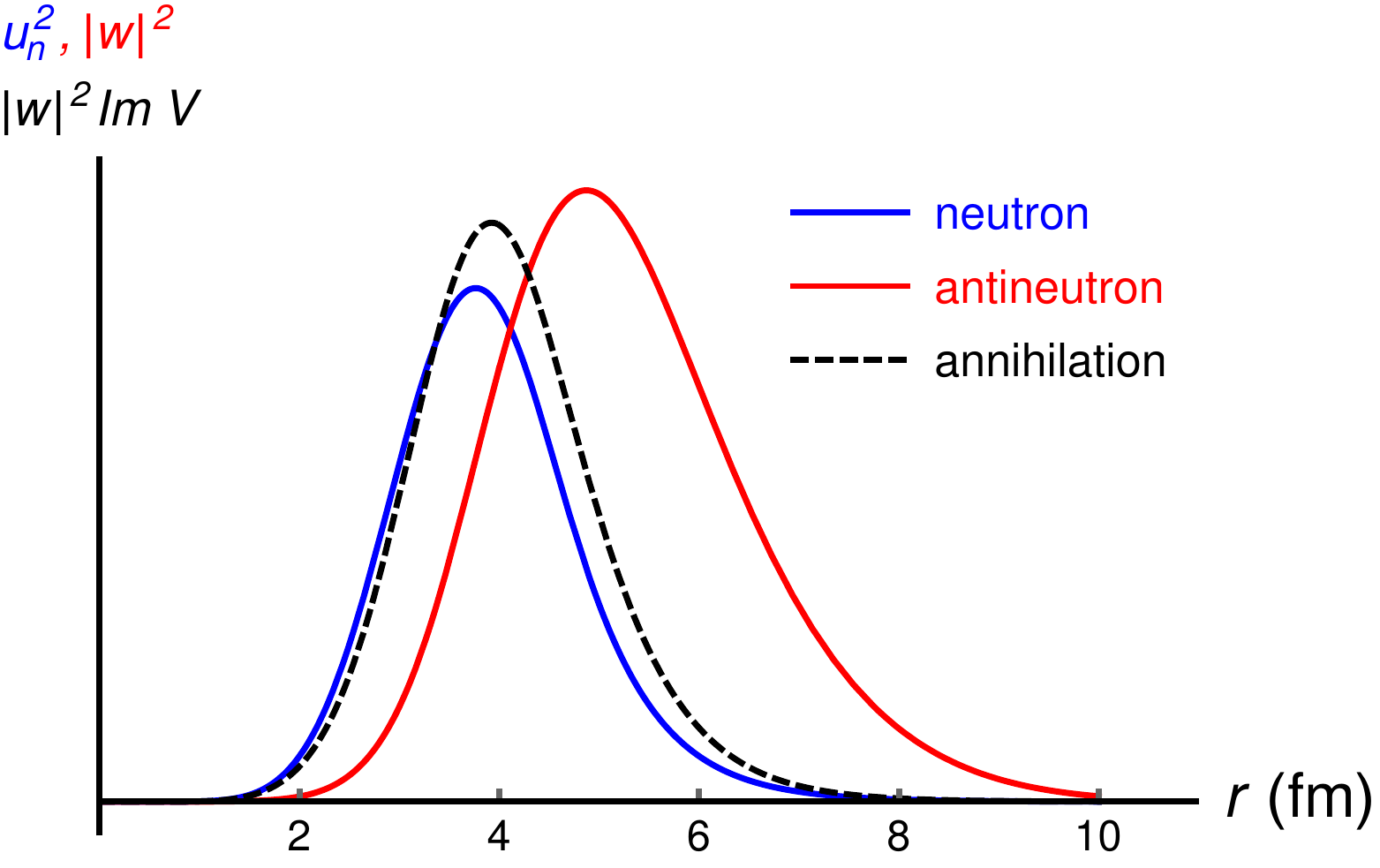}}
 \caption{Same as for Fig.~\ref{fig:Ar1d52}, but for the $1f_{7/2}$ shell.}
\label{fig:Ar1f72}
\end{figure}

\begin{figure}[ht!]
 \centering
 \includegraphics[width=1.0\columnwidth]{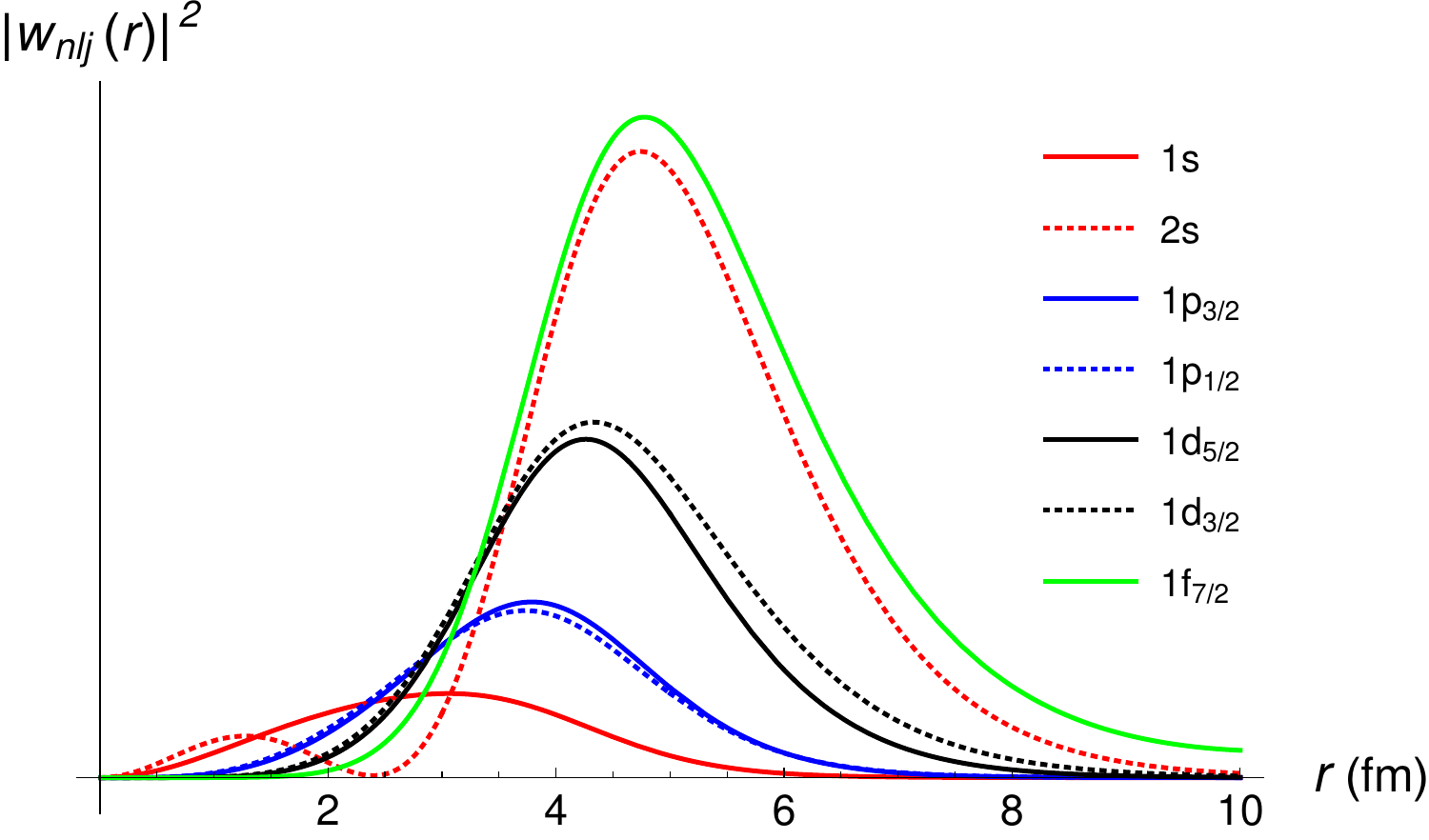}
 \caption{Antineutron densities for the shells of \Arl.}
 \label{fig:Ar1}
\end{figure}

For comparison, the distribution of the $1f_{7/2}$ shell are shown in Fig.~\ref{fig:Ar1f72}: the peripheral character of the antineutron component is even more pronounced, though still decays exponentially as in all other shells.\footnote{Some questions have arisen from colleagues regarding the possibility of an annihilation occurring on \textit{another} nucleus within the larger (detector) medium; this is not permitted due to the exponential decay of the $\bar{n}$ density distribution over \textit{inter}nuclear distances, just as a $n$ is not found inside another nucleus.}
The resulting radial $\bar{n}$ distributions for all shells are shown in Fig.~\ref{fig:Ar1}.

If one now adds up the contributions of each neutron to the width, calculates the average width per $n$, and estimates the corresponding reduced lifetime, one gets a value of $T_R^{Ar}\sim 5.6\times 10^{22}\,\mathrm{s}^{-1}$. As in the case of the deuteron, this value is remarkably stable against changes in the parameters of the $\bar{n}$-nucleon interaction. Thus, a similar uncertainty of $\sim20\,\%$ is estimated.

\begin{figure}[ht!]
 \centering
 \includegraphics[width=1.0\columnwidth]{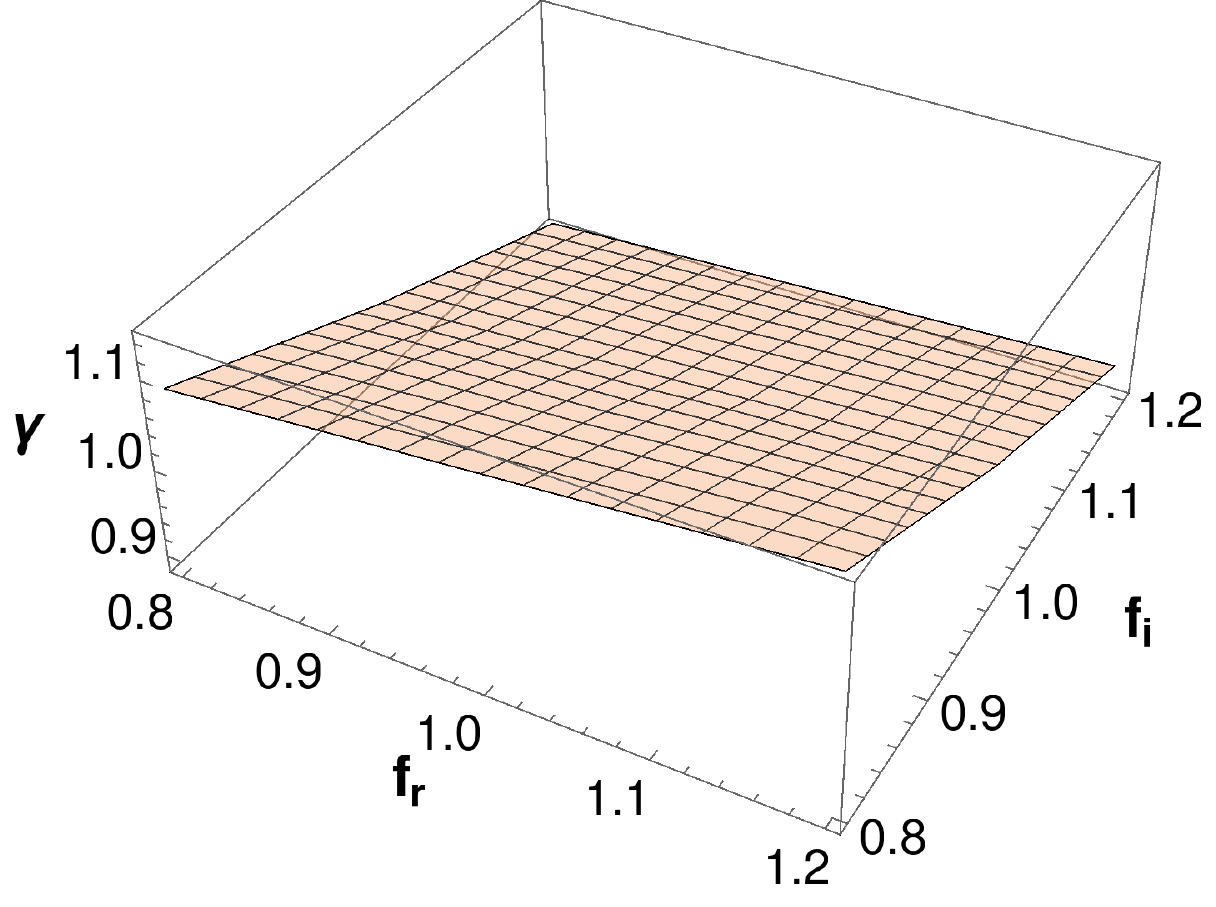}
 \caption{The relative change to the width of the $1d_{5/2}$ shell is shown when a factor $f_r$ is applied to the real potential, and $f_i$ to the imaginary part.}
 \label{fig:stab1}
\end{figure}

This stability in the width can be understood: if one increases the absorptive potential, the $n\to \bar n$ transition is more suppressed, but, on the other hand, the antineutron annihilates more efficiently. In Fig.~\ref{fig:stab1} the factor $\gamma$ is shown, by which the width of the $1d_{5/2}$ level is modified when the real $\bar n$-nucleus potential is multiplied by $f_r$ and its imaginary part by $f_i$. If one changes these values by $\pm20\%$, far beyond what can be admitted to keep a good fit to the antinucleon-nucleus data, modifications to the width are less than $10\%$. The same pattern is observed for the other levels. A consequence of this stability is that the estimated suppression factor $T_R$ keeps the same order of magnitude from one nucleus to the other, even from the deuteron to \Arl.

\begin{figure}[ht!]
 \centering
 \includegraphics[width=1.0\columnwidth]{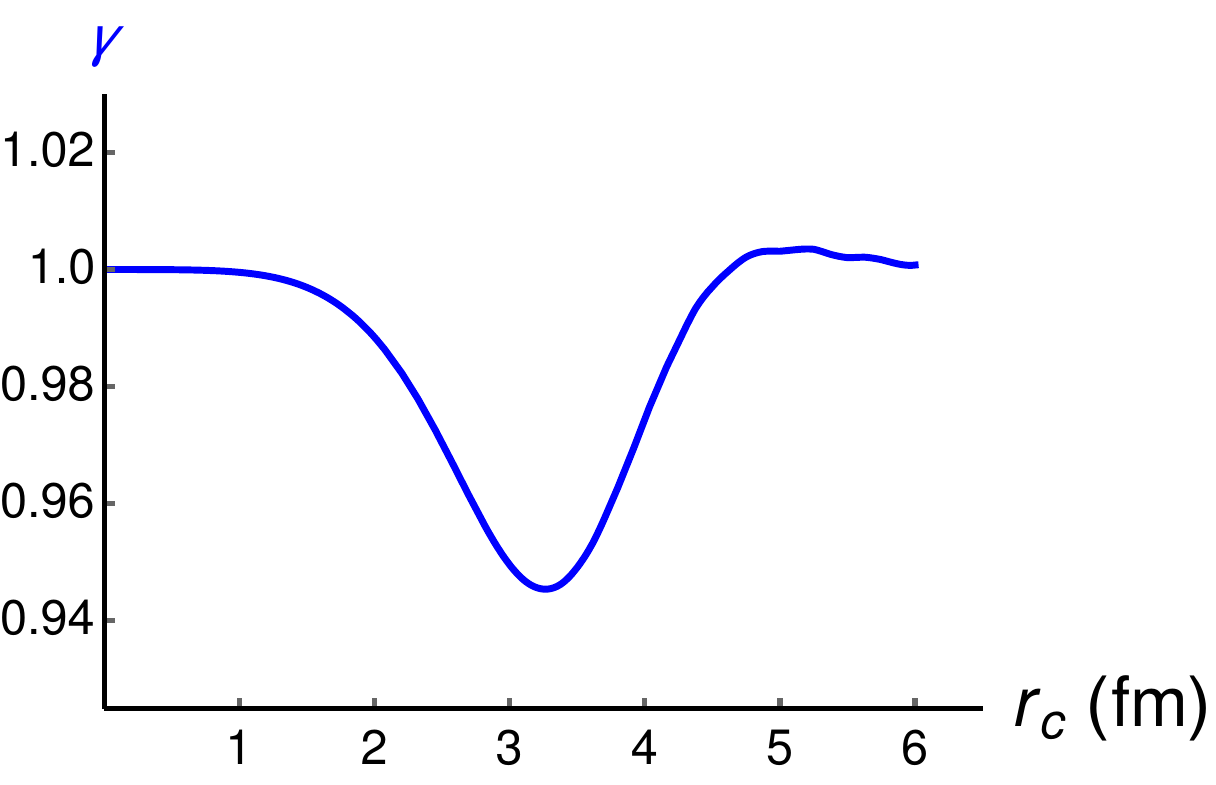}
 \caption{The factor $\gamma$ multiplying the width of the $1d_{5/2}$ shell when a factor $1+0.4\,\exp(-20 (r-r_c))$ ($r$ is in fm) is applied to the absorptive potential.}
 \label{fig:stab2}
\end{figure}

One can also illustrate how intranuclear $n\rightarrow\bar{n}$ transformation and subsequent annihilation is a surface phenomenon. In Fig.~\ref{fig:stab2}, the absorptive potential is modified near $r=r_c$ by applying as a factor a ``glitch'' function with the form $1+0.4 \exp(-20\,(r-r_c)^2)$, where $r_c$ is varied. The width is seen to be modified only near the surface, and is insensitive to what happens in the center of the nucleus. Again, the example shown is  $1d_{5/2}$, but the other levels  exhibit a similar behavior. 

To end this section, it is instructive to compare the method based on an exact solution of the perturbation equation \eqref{eq:Stern-deut} in the 
approximation where the $n$ and $\bar{n}$ spatial distributions are not distinguished. If, furthermore, the $n$- and $\bar n$-nucleus do not differ too much in their real part, then the width is given by (see, e.g., \citep{Chetyrkin:1980ta}):
\begin{equation}
  \label{eq:width-appr}
  \Gamma\approx  -2/\bar W~,\qquad \bar W=\int_0^\infty |u_n(r)|^2 \,W(r)\,\mathrm{d}r~.
\end{equation}
For the low-lying shells of \isotope[40]{Ar}, the difference is small with respect to our estimate. For the most external shell, the width is \textit{overestimated} by a factor of about $3.5$, overemphasizing the suppression of the process.

\boldmath\section{\label{sec:dynamics}Intranuclear $\bar{n}$ Dynamics}\unboldmath
\subsection{\label{sec:Needs}The need for a more precise accounting of nuclear effects}

As a rule, the Intranuclear Cascade (INC) model used for simulations of inelastic interactions of particles within the nucleus is assumed to hold at energies $\gtrsim 30\mbox{-}40$\,MeV (the conditions of applicability of the model are considered in detail in \citep{Golubeva:2018mrz}). The influence of other intranuclear nucleons on an incident particle is taken into account by the introduction of some averaged potential $U(r)$, and so within the nucleus a cascade particle changes its energy by the amount of this potential. In the local Fermi gas approximation used within the INC model, the intranuclear nucleons are bound in the nucleus by a nuclear potential $V_N(r)=-T_{F}(r)-\epsilon_N$, where $T_F$ is the Fermi energy of the nucleon, and $\epsilon_N$ the average binding energy per nucleon. Since the energies of the particles participating in the cascade are sufficiently large, a simplified account of the influence of the nuclear environment is justified and does not lead to distortion of the simulation results

However, after modeling the interaction of a very slow $\bar{n}$ with a $\isotope[12]{C}$ nucleus \citep{Golubeva:2018mrz}, some questions arose about the legitimacy of particular physical approximations which could no longer be ignored, requiring that a more correct accounting of the nuclear environment be attained in both cases of an extranuclear and intranuclear transformation and subsequent annihilation. We now discuss these pertinent changes in our model, which have been incorporated into all extranuclear and intranuclear simulations.

\subsection{\label{sec:modificationofnucelarmedium}The \boldmath{$\bar{n}$}-nucleus intranuclear potential and how \boldmath{$\bar{n}$} modifies the nuclear medium}
The influence of the nuclear environment on an incoming extranuclear (in the case of ESS using \isotope[12]{C} \citep{Golubeva:2018mrz}) \textit{or intranuclear} ${\bar n}$ leads to the modification of the $\bar{n}$'s vacuum four-momentum ${\tilde p}_{\bar n}=(E_{\bar n},{\bf p}_{\bar n})$ (${\tilde p}^2_{\bar n}=E^2_{\bar n}-{\bf p}^2_{\bar n}=m_{n}^2=m_{\bar{n}}^2$) \textit{inside} the target nucleus due to an effective scalar \textit{attractive} nuclear potential of the form \citep{Cassing NPA700}

\begin{equation}
U_{\bar n}(r) = V_0 \, \frac{\rho(r)}{\rho_{0}},\label{eq:potentialeffective}
\end{equation}
where $\rho(r)$ the local nucleon density normalized to the atomic mass number $A$ of the nucleus, $\rho_0$ is the saturation density, $V_0$ is the ${\bar n}$ potential depth at this density, and $r$ is the distance between the ${\bar n}$ and the center of the nucleus.

Using inferences from Fig. \ref{fig:stab1}, one may \textit{assume} that the $\bar{n}A$ and $\bar{p}A$ nuclear potentials are effectively the same. With this potential, a parameter of the model, the total ${\bar n}$ energy $E_{\bar n}^{\prime}$ in the nuclear interior of ordinary nuclei can be expressed in terms of its in-medium mass $m^*_{\bar n}$, defined as \citep{Cassing NPA700,Sibirtsev NPA632} \footnote{$^)$The potential $U_{\bar{n}}$ is the effective $\bar{n}A$ scalar potential. The value of this potential in the center of the nucleus, $V_0$, is actually a free parameter of the model. This determines the total \textit{in-medium} $\bar{n}$ energy through the ``free space" dispersion relation \ref{eq:energybarn} and is not the usual Lorentz scalar potential $U_S^{\bar N}$, determining along with the Lorentz vector potential $U_V^{\bar N}$, the total in-medium antinucleon energy $E_{\bar N}^{\prime}$ via the dispersion relation $E_{\bar N}^{\prime}=\sqrt{(m_N-U_S^{\bar N})^2+{\bf p}_{\bar N}^{{\prime}2}}-U_V^{\bar N}$ \protect\citep{Mishustin PRC71}.}$^)$
%
\begin{equation}
m^*_{\bar n}(r) = m_{\bar{n}} + U_{\bar n}(r)\label{eq:mass_barn}
\end{equation}
and its in-medium three-momentum ${\bf p}_{\bar n}^{\prime}$, as in the free particle case, is \citep{Sibirtsev NPA632}
%
\begin{equation}
E^{\prime}_{\bar n}=\sqrt{m^{*2}_{\bar n}+{\bf p}_{\bar n}^{{\prime}2}}.\label{eq:energybarn}
\end{equation}

Analysis \citep{Sibirtsev NPA632} of $\bar{p}$ production in proton--nucleus and nucleus--nucleus collisions at kinetic energies of a several GeV showed that the $\bar{p}$ potential at normal nuclear matter density is in the range of $-100$ to $-150$\,MeV for outgoing $\bar{p}$ momenta below 2.5\,GeV$/c$. Studies of $\bar{p}$ production at AGS energies \citep{Koch PLB256, Spieles PRC53} suggest ${\bar p}$ potentials of $\simeq-250$ MeV and $\simeq-170$\,MeV at density $\rho_0$ for $\bar{p}$ annihilation events at rest with respect to the nuclear matter and for ${\bar p}$ with momentum of 1\,GeV$/c$, respectively. The real parts of an $\bar{p}$ optical potential in the center of the nucleus of $-(150\pm30)$\,MeV and of $-(220\pm70)$\,MeV were extracted in \citep{Larionov PRC80} from the data on ${\bar p}$ absorption cross sections on nuclei and on the annihilation spectra of $\pi^+$'s and $p$'s, correspondingly. Combined analysis \citep{Friedman NPA761} of data on antiprotonic $X$-rays and of radiochemical data showed that at the center of the nucleus the ${\bar p}$ potential is approximately $-110$\,MeV in depth. So, in spite of various attempts to fix this potential, its depth at density $\rho_0$ \textit{still} remains rather uncertain presently. For the sake of definiteness, in the subsequent calculations, a realistic value of $V_0=-150$\,MeV will be used within Eq.~\eqref{eq:potentialeffective}.

The in-medium momentum ${\bf p}_{\bar n}^{\prime}$ is related to the vacuum momentum ${\bf p}_{\bar n}$ by the following expression:
%
\begin{equation}
E^{\prime}_{\bar n}=\sqrt{m^{*2}_{\bar n}+{\bf p}_{\bar n}^{{\prime}2}}=\sqrt{m_{\bar n}^2+{\bf p}_{\bar n}^{2}}=E_{\bar n}.\label{eq:energyconservation}
\end{equation}

For example, with $V_0=-150$\,MeV, this shows that for ${\bar n}$ annihilation at rest, i.e. when ${\bf p}_{\bar n}=0$, the $\bar{n}$ momentum $|{\bf p}_{\bar n}^{\prime}|$ in the center of the nucleus and at its periphery, corresponding to $10\%$ of the central density, is equal to $510$ and $167$ MeV/c, respectively.

Within the non-interacting local Fermi gas model\footnote{The usual model of a degenerate Fermi gas is one of free nucleons enclosed within a spherical potential well with a sharp border. The intranuclear nucleons fill all energy levels of this well and have momentum from $0$ to the boundary value $P_{FN}$, which is itself a function of the nuclear density. In this case, the momentum of the nucleon does not depend on its position in the nucleus, and so we are thus considering a nonlocal (degenerate) Fermi gas model. Usually, in order to take better account of the influence of the diffuse nuclear boundary, the nucleus is divided into concentric layers (zones), each with a constant density and each with boundary values $P_{FN}^{~i}$. Thus, there now exists a correlation between the position of the nucleon in the nucleus and its momentum, and so we are now considering a local or zoned (degenerate) Fermi gas model. Ample description of this is contained in \citep{Golubeva:2018mrz}.} used in our MC simulations \citep{Golubeva:2018mrz}, for the bound target nucleon total energy $E^{\prime}_{N}$ in the medium at the annihilation point $r$  the formula~\citep{Magas PRC71,Efremov EPJA1}
%
\begin{equation}
E^{\prime}_{N}=\sqrt{m^{2}_{N}+{\bf p}_{N}^{{\prime}2}}+V_N(r)\approx
m_{N}+\frac{{\bf p}_{N}^{{\prime}2}}{2m_N}+V_N(r),\label{eq:energynucleon}
\end{equation}
is used, where for every $i$th concentric zone of a spherical nucleus we have
%
\begin{equation}
V_N^{i}(r)=-\frac{P_{FN}^{~i}{}^2(r)}{2m_N}-\epsilon_N,\label{eq:potentialnucleon}
\end{equation}
with
%
\begin{equation}
P_{FN}^{~i}(r)=\left[3{\pi^2}\rho(r)/2\right]^{1/3}.\label{eq:fermimomentum}
\end{equation}
Here, ${\bf p}_{N}^{\prime}$ is the momentum of the nucleon $N$ ($N=\{p,n\}$) in the Fermi sea, $P_{FN}(r)$ is the boundary Fermi momentum at the local point $r$, and the quantity $\epsilon_N \approx 7$\,MeV is the average binding energy per nucleon; at this point, $0 \le |{\bf p}_{N}^{\prime}| \le P_{FN}(r)$. Within the representation of Eqs.~\eqref{eq:energybarn}--\eqref{eq:energynucleon}, the invariant collision energy $s$ for the interaction of an $\bar{n}$ with a nucleon bound in the nucleus at the point $r$ is
%
\begin{equation}
s=(E^{\prime}_{\bar n}+E^{\prime}_{N})^2-({\bf p}_{\bar n}^{\prime}+{\bf p}_{N}^{\prime})^2=
(E_{\bar n}+E^{\prime}_{N})^2-({\bf p}_{\bar n}^{\prime}+{\bf p}_{N}^{\prime})^2.
\label{eq:energyinvariant}
\end{equation}

The total collision energy $E_{\bar n}+E^{\prime}_{N}$ entering into the second relation of Eq.~\eqref{eq:energyinvariant} in the non-relativistic limit appropriate to our case can be calculated as

\begin{equation}
E_{\bar n}+E^{\prime}_{N}=m_{\bar n}+m_N+\frac{{\bf p}_{\bar n}^{2}}{2m_{\bar n}}+
\frac{{\bf p}_{N}^{{\prime}2}}{2m_N}+V_N(r).\label{eq:energytotal}
\end{equation}

Contrary to the on-shell interaction, for ${\bar n}$ annihilation at rest, from Eq.~\eqref{eq:energytotal} it is seen that this energy is always less than $m_n+m_N$ and its maximum value is $m_n+m_N-\epsilon_N$.

If a bound target $n$ is transformed into an $\bar{n}$ (for the \Arl{}~case), we assume that its total energy $E_n^{\prime}$, defined by Eq.~\eqref{eq:energynucleon}, is equal to that $E_{\bar n}^{\prime}$ of the $\bar{n}$, determined by Eq.~\eqref{eq:energybarn} above. Namely:
%
\begin{equation}
E^{\prime}_{n}=\sqrt{m^{2}_{n}+{\bf p}_{n}^{{\prime}2}}+V_N(r)=
\sqrt{m^{*2}_{\bar n}+{\bf p}_{\bar n}^{{\prime}2}}=E^{\prime}_{\bar n}.
\label{eq:energyrelation}
\end{equation}

It is interesting to note that, for $p_{F}(0)=250$\,MeV$/c$, Eq.~\eqref{eq:energyrelation} gives for the ${\bar n}$ momentum $|{\bf p}_{\bar n}^{\prime}|$ the values of 430 and 40\,MeV$/c$ if the transition $n \to {\bar n}$ of the target $n$ at rest occurs in the center of the nucleus and at its periphery, respectively (corresponding to $10\%$ of the central density).

The value of $s$ for the intranuclear \Arl{}~case is given by the first relation of formula \eqref{eq:energyinvariant}.

\section{\label{sec:MonteCarlo}Monte Carlo Simulation Methods}

In the preceding sections, we have attempted to summarize new theoretical additions to the dynamics of our MC simulation. We will now review the stages of our generator, accompanied by some discussion of other recent changes given new antinucleon annihilation data, and show some differences from previously published outputs \citep{Golubeva:2018mrz}.

\subsection{\label{sec:fundamentals}Fundamentals}

Most all of the background concerning our MC generator can be found in our recent work \citep{Golubeva:2018mrz}, however, we will mention some improvements which have been implemented here. We can briefly summarize the content of the model as follows:

\begin{enumerate}
    \item Unlike the optical approach used in our previous work \citep{Golubeva:2018mrz} to calculate the radial annihilation position distribution of a slow external antineutron annihilating on \isotope[12]{C}, in this work, an intranuclear annihilation point is taken from a corresponding probability distribution shown later for \Arl~in Fig.~\ref{fig:AnnDist}. This point lies within a particular \textit{zone} of a set of eight concentric spherical shells representing the volume of the nucleus. Each shell has its own uniform nuclear density and each their own single nucleon momentum distribution, thus acting as a \textit{zoned} local Fermi gas.
    \item The annihilation occurs, producing $\pi$'s and higher mass resonances such as $\eta,\omega$, and $\rho$'s, according to tabulated channels with  branching ratios taken from \citep{Golubeva:2018mrz}. After the decay of all resonances, on average $\sim 4\mbox{-}5$ mesons are produced from the initial annihilation process.
    \item The annihilation products are then transported through the nuclear environment quasi-classically using a full intranuclear cascade model (a local nuclear density decrease is also included). Meson resonances are decayed according to their individual lifetimes inside and outside the nucleus, though particularly long-lived species are treated as stable inside the nucleus.
    \item The products are ejected from the nucleus and the nuclear remnant(s) is allowed to de-excite, evaporating nucleons and fragments of higher mass.
\end{enumerate}

\subsection{\label{sec:newdata}Improvements from OBELIX and Crystal Barrel Data}

The annihilation model was created in 1992 \citep{Golubeva:1992tr}, originally using experimental branching ratios obtained before this time. Later analysis of experimental data from $\bar{p}p$ annihilation at rest obtained from the LEAR (CERN) $\bar{p}$ beam by the OBELIX \citep{Salvini:2004gz} and Crystal Barrel \citep{Amsler:2003bq} collaborations are now integrated into the internal annihilation model. Note that all tables in all proceeding sections are outputted from samples of 10,000 events. In Table~\ref{tab:newbranchingfractions} we see the absolute changes in the branching fractions of individual annihilation channels in accordance with this newer experimental data. The first column shows the annihilation channel, the second column shows the value of this channel's branching fraction in the corresponding table in \citep{Golubeva:2018mrz}, while the third shows the new branching fraction of this channel changed in accordance with the experimental data \citep{Salvini:2004gz,Amsler:2003bq}. Once summed, all channels are then renormalized to unity with these new fractions taken into account. A comparison of the elementary processes for $\bar{p}p$ annihilation was carried out taking into account experimental data \citep{Salvini:2004gz,Amsler:2003bq} and annihilation simulations from our recent work \citep{Golubeva:2018mrz}.

\onecolumngrid

\begin{center}
\begin{table}[ht!]
 \caption{A list of several $\bar{p}p$ annihilation products and their respective old \protect\citep{Golubeva:2018mrz} and new branching fractions with inclusion of data from \protect\citep{Salvini:2004gz,Amsler:2003bq}.}
    \label{tab:newbranchingfractions}
\begin{ruledtabular}
    \begin{tabular}{c|cc}
        Channel & Probability (\%) from \protect\citep{Golubeva:2018mrz} & Probability (\%) Used for This Work \\ \hline
        $\bar{p}p\rightarrow\pi^+\pi^+\pi^-\pi^-$ & $2.74$ & $3.64$~\protect\citep{Salvini:2004gz} \footnote{OBELIX gives data for $\bar{p}p\rightarrow\pi^+\pi^+\pi^-\pi^-$ with a branching ratio of $BR(\bar{p}p\rightarrow\pi^+\pi^+\pi^-\pi^-)=6.4\pm0.09\%$ (a gas target). This value includes both \textit{resonant} and independent production of these four charged $\pi$'s. To avoid double counting, we have subtracted from this value all the fractions of all of the channels that give rise to $\pi^+\pi^+\pi^-\pi^-$ in the final state. Thus, in the $\bar{p}p$ annihilation table contained in \protect\citep{Golubeva:2018mrz}, we have \textit{experimental} data showing $BR(\bar{p}p\rightarrow\rho^0\rho^0)=0.67\%$, $BR(\bar{p}p\rightarrow\pi^+\pi^-\rho^0)=2.02\%$, and importantly $BR(\bar{p}p\rightarrow\pi^+\pi^-\omega)=3.03\%$, whose contribution to the full channel of $\bar{p}p\rightarrow\pi^+\pi^+\pi^-\pi^-$ (including $BR(\omega\rightarrow\pi^+\pi^-)=2.3\%$ is $3.03\times0.023=0.07$. From all of these considerations, we introduce the final ratio to the table as $6.4-0.67-2.02-0.07=3.64\%$.}\\ 
        $\bar{p}p\rightarrow\pi^0\pi^0$ & $0.02$ & $0.154$~\protect\citep{Amsler:2003bq}\\
        $\bar{p}p\rightarrow\eta\eta$ & $0.01$ & $0.0312$~\protect\citep{Amsler:2003bq}\\ 
        $\bar{p}p\rightarrow\pi^0\omega$ & $0.58$ & $0.460$~\protect\citep{Amsler:2003bq}\\ 
        $\bar{p}p\rightarrow\eta\omega$ & $0.34$ & $0.960$~\protect\citep{Amsler:2003bq}\\ 
        $\bar{p}p\rightarrow\pi^0\pi^0\pi^0$ & $1.12$ & $0.610$~\protect\citep{Amsler:2003bq}\\ 
        $\bar{p}p\rightarrow\pi^0\pi^0\eta$ & $0.54$ & $0.514$~\protect\citep{Amsler:2003bq}\\
        $\bar{p}p\rightarrow\omega\omega$ & $1.57$ & $0.358$~\protect\citep{Amsler:2003bq}\\ 
        $\bar{p}p\rightarrow\pi^0\pi^0\omega$ & $0.79$ & $1.53$~\protect\citep{Amsler:2003bq}\\ 
        $\bar{p}p\rightarrow\eta\pi^0\omega$ & $0.30$ & $ 0.60$~\protect\citep{Amsler:2003bq}\\
        $\bar{p}p\rightarrow\pi^0\omega\omega$ & $0.37$ & $0.344$~\protect\citep{Amsler:2003bq}\\ 
        $\bar{p}p\rightarrow\pi^0\pi^0\pi^0\omega$ & $0.40$ & $1.24$~\protect\citep{Amsler:2003bq}\\ 
    \end{tabular}
    \end{ruledtabular}
\end{table}
\end{center}

\twocolumngrid
Table \ref{tab:newmesonmultiplicities} shows the average multiplicities of mesons produced in $\bar{p}p$ annihilation at rest. The first column re-presents the simulation results from Table IV in \citep{Golubeva:2018mrz}, while the second column presents the results of modeling when taking into account data from \citep{Salvini:2004gz,Amsler:2003bq}. The third column also re-presents the experimental data itself \citep{Klempt:2005pp,Minor,Levman:1979gg,Hamatsu:1976qz,Golubeva:2018mrz} for ease of comparison. It follows from Table \ref{tab:newmesonmultiplicities} that the average multiplicities of annihilation mesons with changes in the branching ratios of some individual channels do not change significantly and the difference between the two calculation options is within the uncertainty of the experimental data. Nevertheless, in this and all future simulations, this new table will be used for modeling all $\bar{p}p$ annihilations.

\onecolumngrid

\begin{center}
\begin{table}[ht!]
  \caption{Meson multiplicity comparisons from previous work \protect\citep{Golubeva:2018mrz} and the new generator as described here, which include experimental integration of more recent OBELIX \protect\citep{Salvini:2004gz} and Crystal Barrel \protect\citep{Amsler:2003bq} data sets.}
    \label{tab:newmesonmultiplicities}
\begin{ruledtabular}
    \begin{tabular}{c|ccc}
        & $\bar{p}p$ Simulation from \protect\citep{Golubeva:2018mrz} & $\bar{p}p$ Simulation with New Model & $\bar{p}p$ Experiment \\ \hline
        $M(\pi)$ & $4.91$ & $4.95$ & $4.98\pm0.35$~\protect\citep{Klempt:2005pp}, $4.94\pm0.14$~\protect\citep{Minor}\\
        $M(\pi^{\pm})$ & $3.11$ & $3.09$ & $3.14\pm0.28$~\protect\citep{Klempt:2005pp}, $3.05\pm0.04$~\protect\citep{Klempt:2005pp}, $3.04\pm0.08$~\protect\citep{Minor}\\
        $M(\pi^0)$ & $1.80$ & $1.86$ & $1.83\pm0.21$~\protect\citep{Klempt:2005pp}, $1.93\pm0.12$~\protect\citep{Klempt:2005pp}, $1.90\pm0.12$~\protect\citep{Minor}\\
        $M(\eta)$ & $0.09$ & $0.09$ & $0.10\pm0.09$~\protect\citep{Levman:1979gg}, $0.07\pm0.01$~\protect\citep{Klempt:2005pp}\\
        $M(\omega)$ & $0.20$ & $0.27$ & $0.28\pm0.16$~\protect\citep{Levman:1979gg}, $0.22\pm0.01$~\protect\citep{Hamatsu:1976qz}\\ 
        $M(\rho^+)$ & $0.19$ & $0.19$ & -----\\ 
        $M(\rho^-)$ & $0.19$ & $0.18$ & -----\\ 
        $M(\rho^0)$ & $0.19$ & $0.18$ & $0.26\pm0.01$~\protect\citep{Hamatsu:1976qz}    \end{tabular}
    \end{ruledtabular}
\end{table}
\end{center}
 
\twocolumngrid

\subsection{\label{sec:newdatacomparison}Model modifications and a new comparison with experimental \boldmath{$\bar{p}C$} annihilation at rest}

The branching fraction modifications shown in Table~\ref{tab:newbranchingfractions} were implemented into the optical-cascade model, most recently discussed in \citep{Golubeva:2018mrz}. Now, let's analyze how these changes effect the description of experimental data for $\bar{p}C$ annihilation at rest. The first line of Table \ref{tab:multiplicities-exper_old_calc1_calc2} presents the experimental multiplicities of the emitted $\pi$'s and the energy carried away by those $\pi$'s and $\gamma$'s. The second line of the table shows the results of a calculation with our original optical-cascade model from \citep{Golubeva:2018mrz} before any modifications. The third line (Calculation $\# 1$) presents the results of a simulation taking into account the changes in the annihilation table described above. The last line (Calculation $\# 2$) presents the results of a simulation accounting for both the changes in the annihilation table and modifications related to the nuclear environment. In both cases, an antinucleon potential of $150$\,MeV in the center of the nucleus, and varying in accordance with the nuclear density, is included.

\onecolumngrid

\begin{center}
\begin{table}[ht!]
\caption{A list of updated multiplicities from experimental data, our original work concerning \isotope[12][6]{C} \protect\citep{Golubeva:2018mrz}, and two new calculations taking into account the newest versions of $\bar{p}$ annihilation branching ratios while also considering a new intranuclear antinucleon potential with an associated nuclear medium response.}
\label{tab:multiplicities-exper_old_calc1_calc2}
\begin{ruledtabular}
    \begin{tabular}{c|ccccccc}
     & $M(\pi)$ & $M(\pi^+)$ & $M(\pi^-)$ & $M(\pi^0)$ & $E_{tot}$\,(MeV) & $M(p)$ & $M(n)$\\ \hline
    $\bar{p}{\rm C}$ Experiment & $4.57 \pm 0.15$ & $1.25 \pm 0.06$ & $1.59 \pm 0.09$ & $1.73 \pm 0.10$ & $1758 \pm 59$ & ----- & -----\\
    $\bar{p}{\rm C}$ Old Calculation & $4.56$ & $1.21$ & $1.63$ & $1.72$ & $1736$ & $1.14$ & $1.21$\\
    $\bar{p}{\rm C}$ Calculation $\# 1$ & $4.56$ & $1.21$ & $1.63$ & $1.72$ & $1738$ & $1.11$ & $1.20$\\
    $\bar{p}{\rm C}$ Calculation $\# 2$ & $4.60$ & $1.22$ & $1.65$ & $1.73$ & $1762$ & $0.96$ & $1.03$\\
    \end{tabular}
\end{ruledtabular}
\end{table}
\end{center}

\twocolumngrid

From analysis of the table, it follows that the average multiplicities vary slightly with the modification of the annihilation table and dynamics. Multiplicities of $\pi$'s increase \textit{slightly} with the modifications associated with the influence of the nuclear environment, and everywhere are within the experimental error. At the same time, the multiplicity of $p$'s and $n$'s emitted during the cascade development and emitted in the process of de-excitation was significantly reduced in the Calculation $\# 2$. This \textit{suggests} that the dynamics of meson-nuclear interactions and energy dissipation in the residual nucleus change somewhat with the introduction of the influence of the environment. Clarification of this issue requires data on proton and neutron emission from $\bar{p}A$ annihilation experiments, along with further detailed study. Since the present work is devoted to experiments to search for transformations that are planned to be registered through the observation of predominately spherically symmetric, multipionic topologies (so-called $\pi$-stars), we will focus on the characteristics of these topologies and their associated quantities.

\begin{figure}[ht!]
    \centering
    \includegraphics[width=1.0\columnwidth]{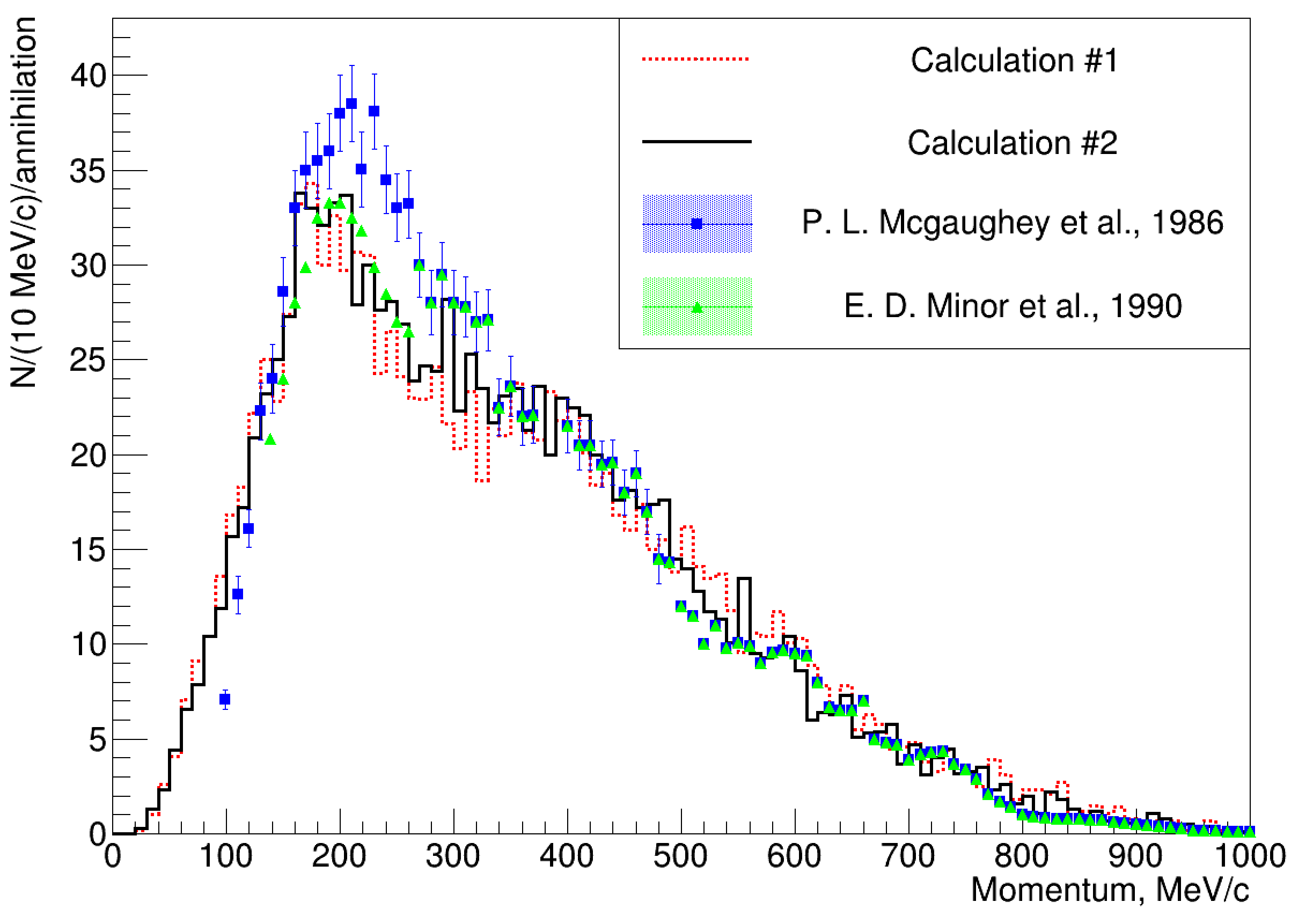}
    \caption{The momentum distribution for $\pi^+$ emitted from $\bar{p}{\rm C}$ annihilation at rest. The dashed red histogram shows the distribution generated from Calculation $\# 1$ mentioned in Table \ref{tab:multiplicities-exper_old_calc1_calc2}, while the solid black shows Calculation $\# 2$. All points are taken from experimental data in \citep{Minor,Mcgaughey:1986kz}.}
    \label{fig:pbarCarbon-PiPlusMomentum}
\end{figure}

Fig.~\ref{fig:pbarCarbon-PiPlusMomentum} shows the spectrum of $\pi^+$ emitted during $\bar{p}{\rm C}$ annihilation at rest in comparison to experimental data from \citep{Minor} (green triangles), and \citep{Mcgaughey:1986kz} (blue squares). Just as in the average multiplicities shown in Table~\ref{tab:multiplicities-exper_old_calc1_calc2}, there are no significant differences in the spectrum for these two variants of calculation, though a comparison to the old calculation reveals a slightly better fit to the data in the region around $250$\,MeV (the $\Delta$-resonance).

In our proceeding simulations for the \Arl{} nucleus to be discussed in later sections, we will use a model analogous to Calculation $\#2$ that includes effects related to the influence of the nuclear environment.

\section{\label{sec:NewResults}New results for simulations of $\bar{n} A$ annihilation}

\subsection{\label{sec:simulationchanges}Changes in extranuclear \boldmath{$\bar{n}C$} annihilation simulations}

\begin{figure}[ht!]
    \centering
    \includegraphics[width=1.0\columnwidth]{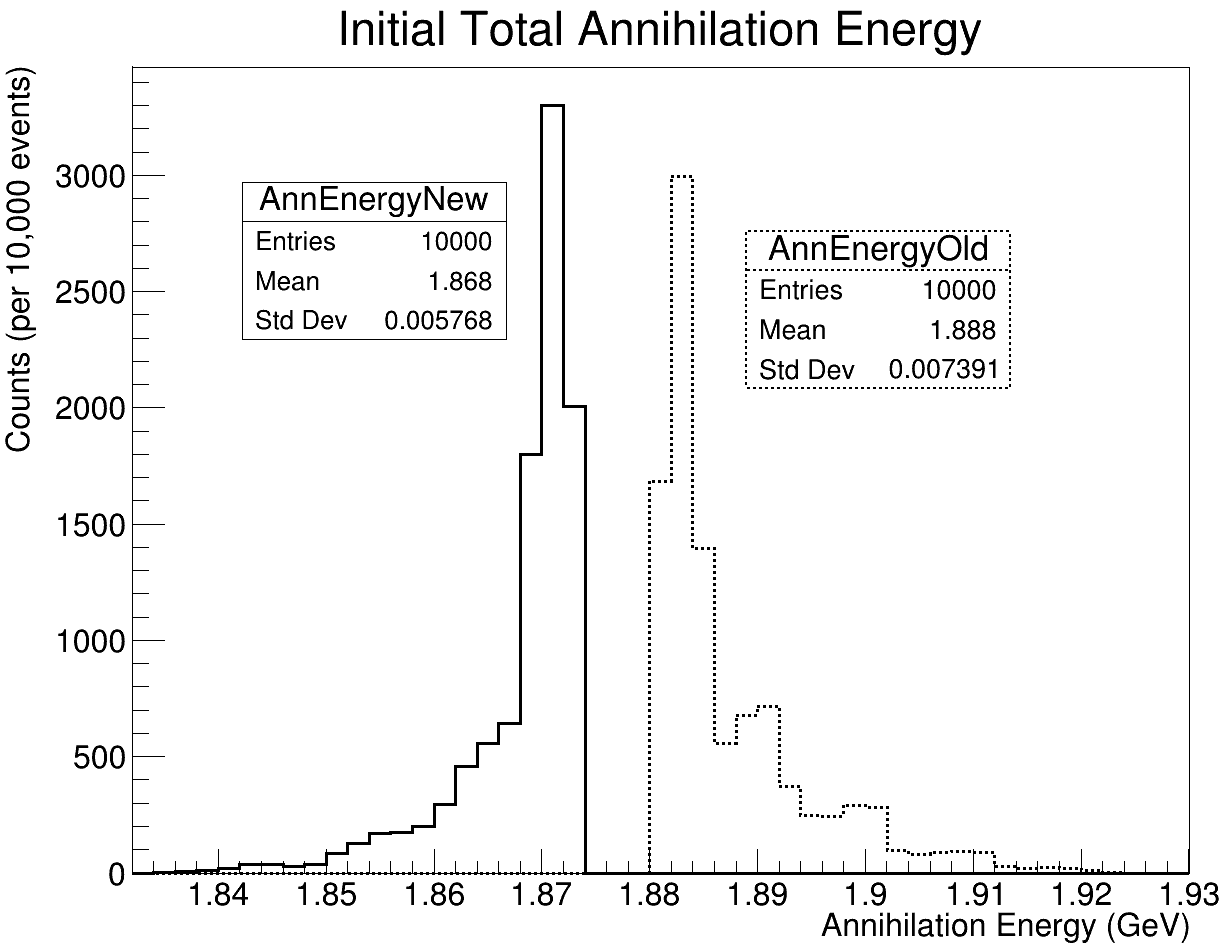}
    \caption{The new (solid line) and old (dashed line) total energy available to annihilating (anti)nucleons (and generated mesons) is plotted.}
    \label{fig:OldNewnbarCInitAnnEnergy}
\end{figure}

\begin{figure}[ht!]
    \centering
    \includegraphics[width=1.0\columnwidth]{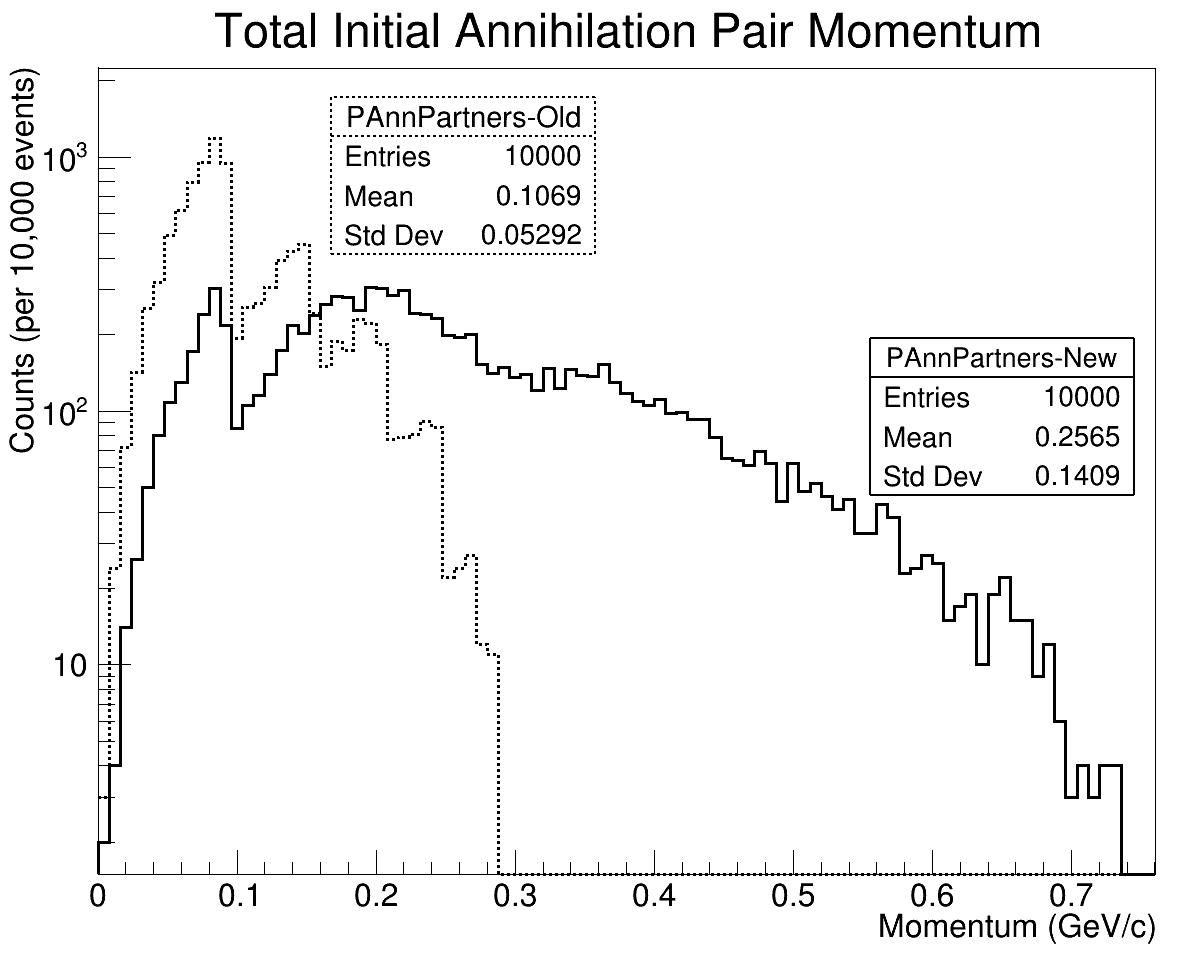}
    \caption{The old (dashed line) and new (solid line) total $\bar{n}N$ pair (or generated meson) momentum is plotted.}
    \label{fig:OldNewInitnbarNPairMom}
\end{figure}

\begin{figure}[ht!]
    \centering
    \includegraphics[width=1.0\columnwidth]{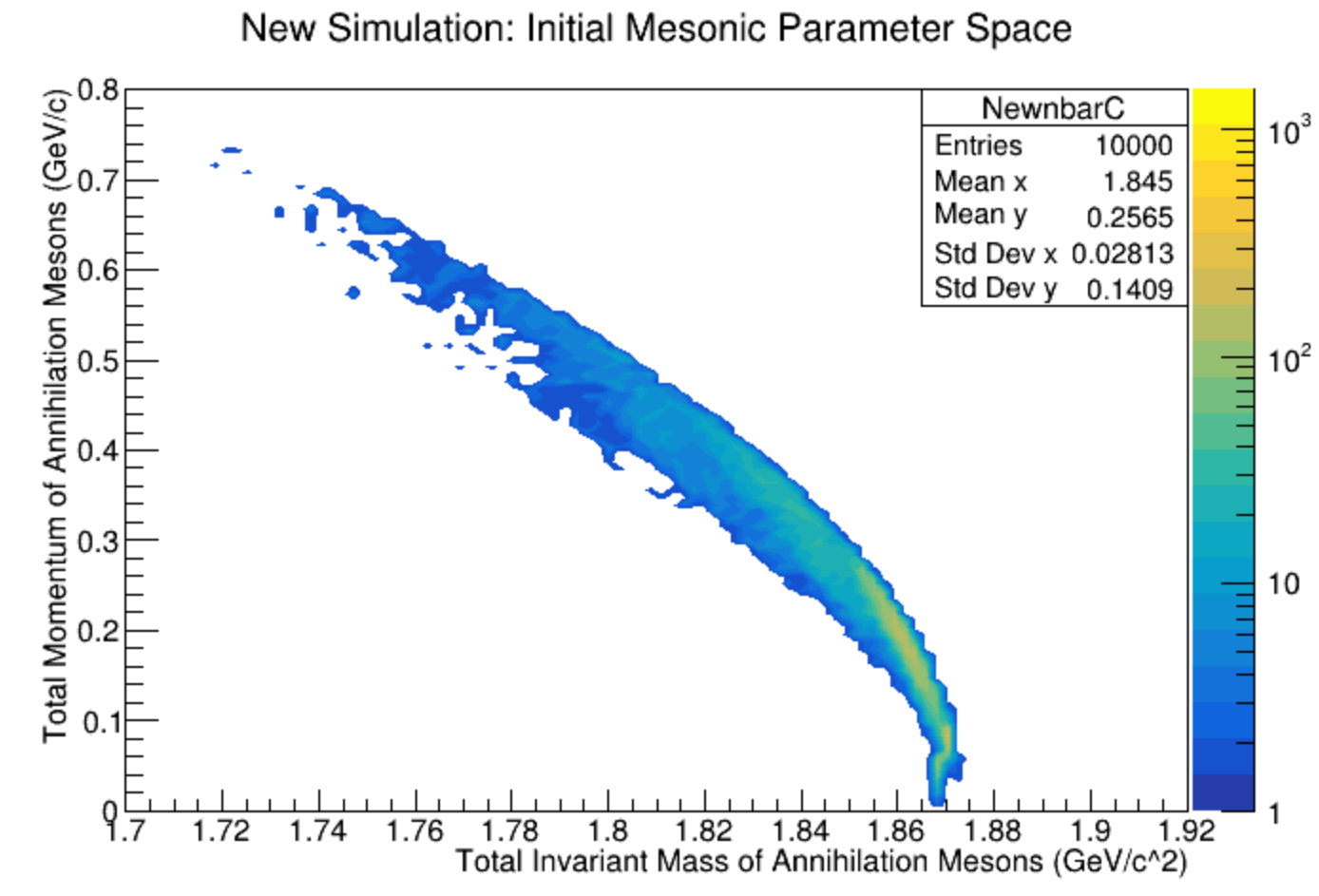}
    \includegraphics[width=1.0\columnwidth]{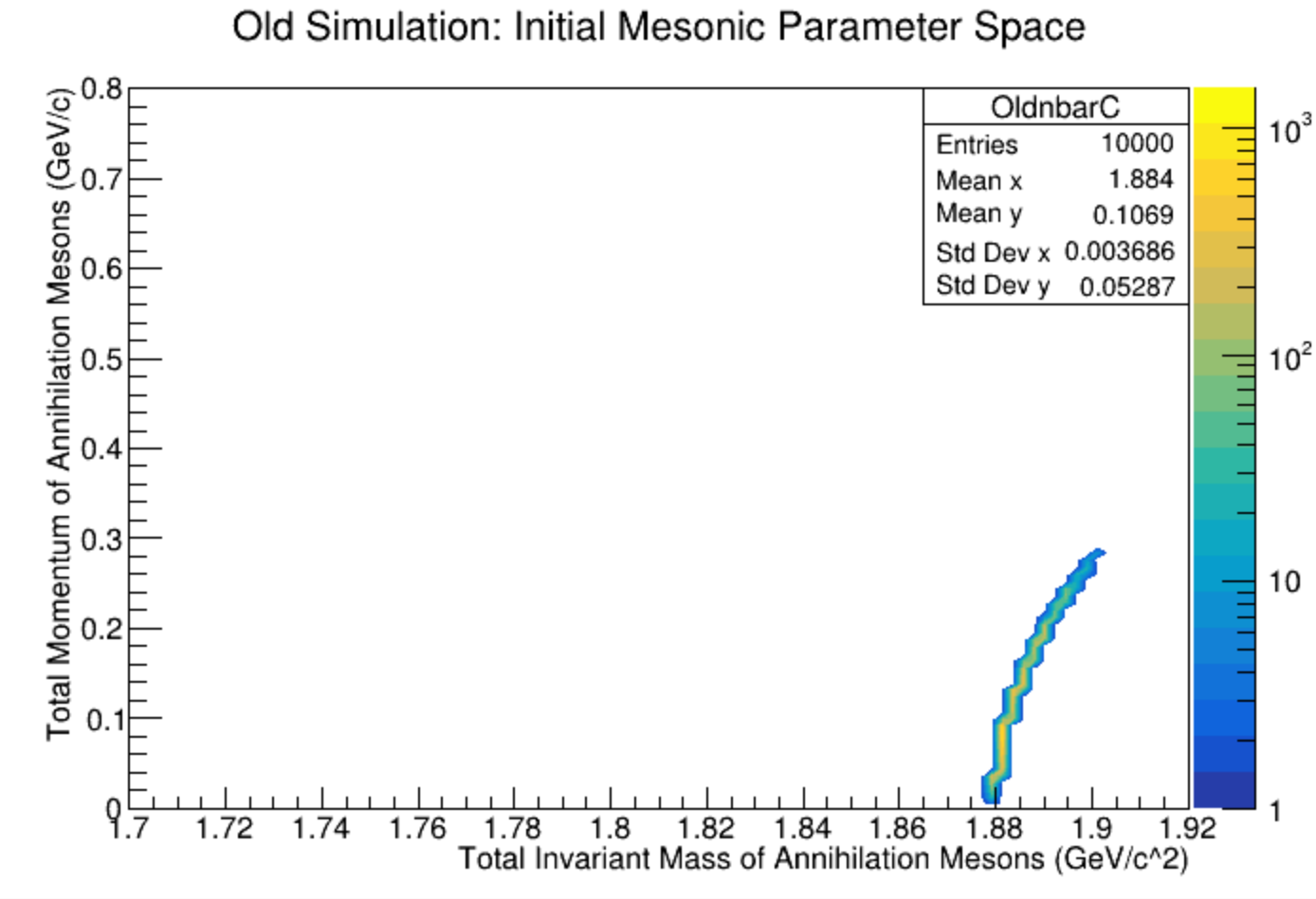}
    \caption{The new (top) and old (bottom) total mesonic initial parameter space is shown for extranuclear $\bar{n} C$ annihilation.}
    \label{fig:OldNewInitMesMomvsInvMass}
\end{figure}

\begin{figure}[ht!]
    \centering
    \includegraphics[width=1.0\columnwidth]{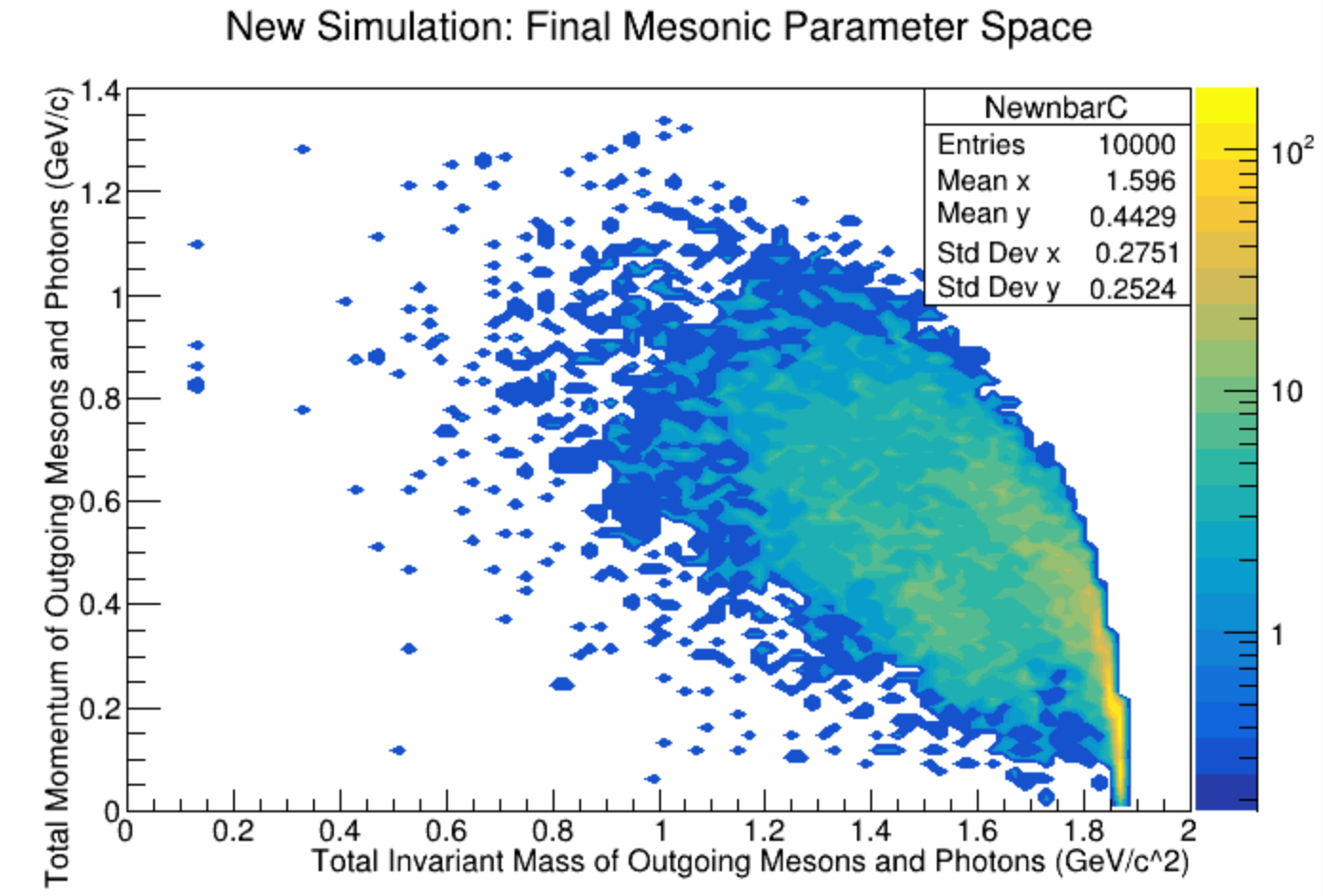}
    \includegraphics[width=1.0\columnwidth]{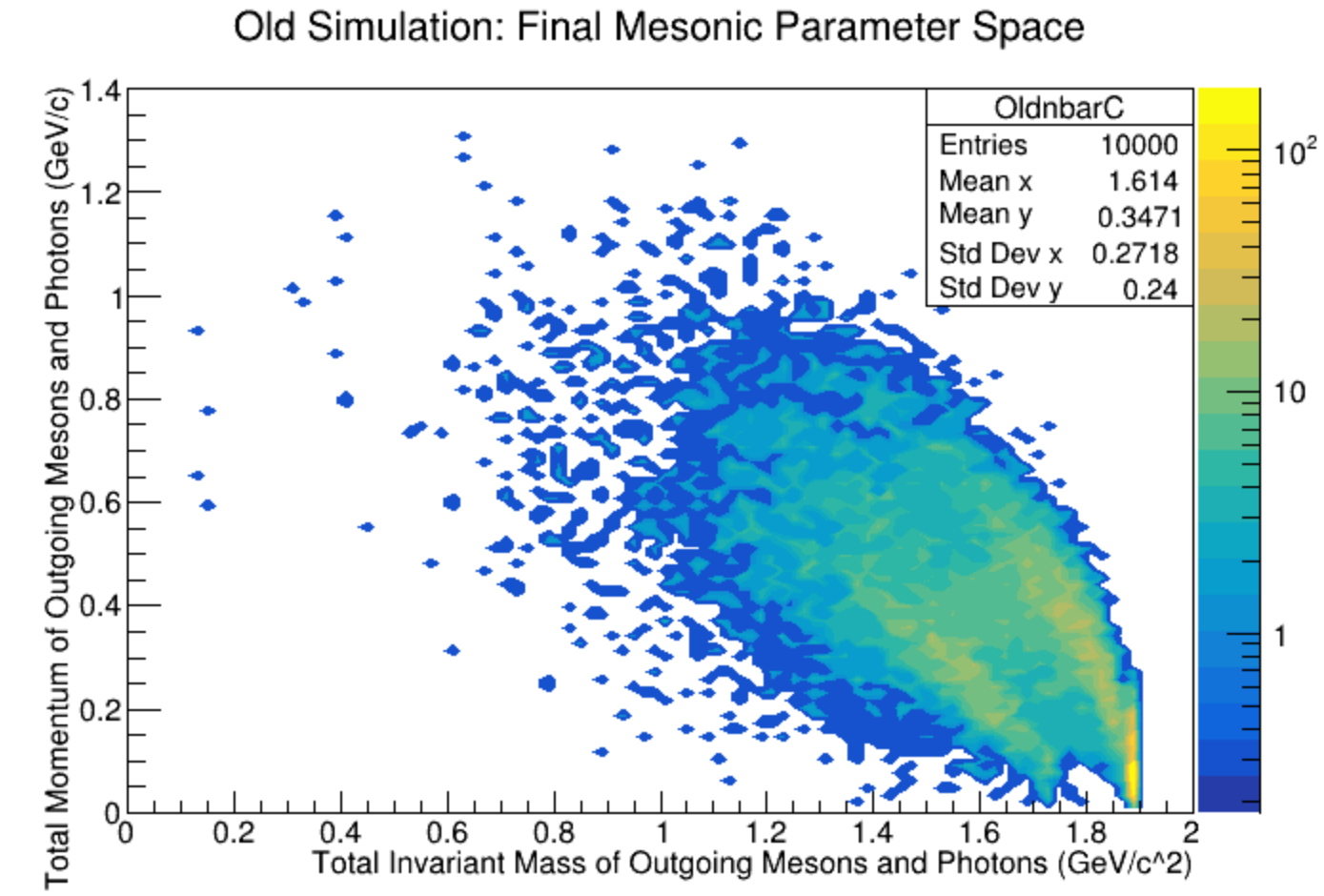}
    \caption{The new (top) and old (bottom) total mesonic/pionic/photonic final state parameter space is shown for extranuclear $\bar{n} C$ annihilation.}
    \label{fig:OldNewFinMesMomvsInvMass}
\end{figure}

Despite the absence of significant differences in the description of available $\bar{p}{\rm C}$ experimental data, there are certain aspects of the two variants of the models where differences are very noticeable. The changes discussed above lead to some rather important differences in annihilation stages as compared to previous work in \citep{Golubeva:2018mrz}, examples of which can be seen in Figs.~\ref{fig:OldNewnbarCInitAnnEnergy}--\ref{fig:OldNewFinMesMomvsInvMass}. These can make possible different observable final states after the intranuclear cascade which are important for experiments planning to utilize smaller nuclei (such as the NNBar Collaboration at the European Spallation Source). Note that all plots in all proceeding sections are outputted from samples of 10,000 events.

For the intranuclear cascade, the operating energy conservation law for the annihilation process of an \textit{extranuclear} neutron is written as
\begin{equation}
    E_{ann} + E^{*} = E^{'}_{\bar{n}} + E^{'}_{N},
\end{equation}
where $E^{'}_{\bar{n}}$ is the total energy of the $\bar{n}$ inside the nucleus at the point of annihilation, $E^{'}_{N}$ is the total energy of the nucleon annihilation partner at the same point, and $E^{*}$ is the excitation energy of the nucleus after the annihilation. In the degenerate Fermi gas model, this is defined as
\begin{equation}
    E^{*} = T^{i}_{FN}-T^{i}_{N},
\end{equation}
varying from $0$ to $T^{i}_{FN}$, where $i$ is the zone number in which the annihilation takes place, $T^{i}_{FN}$ is the boundary Fermi energy of the $i$-th zone, and $T^{i}_{N}$ is the Fermi energy of the annihilation partner in the same zone. If one takes into account relations \eqref{eq:energyconservation}--\eqref{eq:potentialnucleon} and \eqref{eq:energytotal} from Sec.~\ref{sec:dynamics}, it follows that
\begin{equation}
    E_{ann} + E^{*} = m_{\bar{n}} + m_{N} -\epsilon_N,
\end{equation}
where $\epsilon_N = 7$\,MeV/nucleon.

In Fig.~\ref{fig:OldNewnbarCInitAnnEnergy}, we see the distributions (old and new) of the initial amount of total energy carried by the two annihilating nucleons. Due to the fact that the previous iteration of our calculations did not include a true $\bar{n}$-potential or consider intranuclear nucleons to be off-shell, the skew of this distribution was always \textit{greater} than the available rest-mass energy of an annihilating pair; this was discounted as a kind of \textit{virtual} intranuclear effect, which later disappeared and showed conservation of energy and momentum once entering the final state. This is now changed and made internally consistent, showing a proper distribution \textit{less than} the combined free rest masses of the pair of annihilating nucleons. Secondarily, the strength of the $\bar{n}$-potential can be seen to smooth out the zoned structure present in the previous calculation.

The dynamical (position-correlated) momentum of the initial annihilation pair (and, by conservation, their annihilation products) can be studied in Fig.~\ref{fig:OldNewInitnbarNPairMom}. In the old variant of the model, the $\bar{n}$ was assumed to come from a transformation down-range of a cold $n$ source with a mean energy of only $\sim$meV, ignoring the $\bar{n}$-potential. Thus, the original momentum distribution of the annihilation products (dashed line) was effectively a direct observation of the non-interacting zoned local Fermi gas single nucleon momentum distribution folded with the radial annihilation probability distribution. In the new variant of the model described above, shown in the solid histogram, the mass of the $\bar{n}$ is defined by expression \eqref{eq:mass_barn} and the momentum of the $\bar{n}$ follows from \eqref{eq:energyconservation}. The direction of the momentum of the nucleons are isotropically distributed, and thus the total momentum of the annihilation products varies in absolute value from $|P^{'}_{\bar{n}}-P^{'}_{N}|$ to $|P^{'}_{\bar{n}}+P^{'}_{N}|$, smoothing and spreading out the structure associated with the presence of zones in the target nucleus. The peak in the histogram in the region just below $100$\,MeV/$c$ corresponds to annihilations on the outside of the nucleus (within the diffuse eighth zone), where the $\bar{n}$-potential is taken to be $0$ while no off-shell mass is accounted for, and so the momentum of the annihilating pairs is equal to the momentum of the nucleon partner within the seventh zone.

The most impressive new figures in consideration of $\bar{n}{\rm C}$ extranuclear annihilation are most likely Figs.~\ref{fig:OldNewInitMesMomvsInvMass} and \ref{fig:OldNewFinMesMomvsInvMass}. Here we see the total available initial and final mesonic/pionic and photonic parameter space for an $n\rightarrow\bar{n}$ signal, most commonly shown via total momentum versus invariant mass plots (\textit{similar} to Fig. 2 in \citep{Abe15}) at the annihilation point before any nuclear transport is completed. The results of new modeling are shown at top, and the old version of the model is shown below. In Figs.~\ref{fig:OldNewInitMesMomvsInvMass}, the effect of the antineutron potential and off-shell nature of nucleon masses are clearly seen in the top plot, while the bottom plot shows a small momentum range due to the absence of the antineutron momentum; the right-ward inclination of the old parameter space is a consequence of the on-shell mass' effects on the overall kinematics of the annihilation. Thus, we have significantly different \textit{initial} conditions for the transport of annihilation mesons through the nucleus.

Figs.~\ref{fig:OldNewFinMesMomvsInvMass} show the same variables after transport, but re-scattering, $\Delta$-resonance, and absorption of annihilation mesons leads to an overall decrease in the observed invariant mass, lessening the apparent final state differences between the two simulation variants.
Overall, the new parameter space is slightly more spatially if not statistically constrained, which could lead to higher hypothetical experimental efficiencies for future experiments.

\subsection{\label{sec:intranbarargonsim}Intranuclear \boldmath{$\bar{n}{\rm  Ar}$} annihilation simulations}
As mentioned previously, there now exist multiple generators for $n\rightarrow\bar{n}$ within the particle physics community, notably developed in \citep{Hewes:2017xtr} using GENIE \citep{Andreopoulos:2015wxa}. While we believe that the demonstration of our independent generator's capabilities \citep{Golubeva:2018mrz} in the reproduction of antinucleon data is well-established for $\isotope[12]{C}$ \citep{Golubeva:2018mrz}, such a \textit{complete} set of physical observables does not readily exist to constrain the model for larger nuclei, especially not for \Arl. Thus, out of a need for ample comparisons, we endeavor to show the commonalities and differences between each of these $n\rightarrow\bar{n}$ generators for intranuclear $\bar{n}\,\isotope[39]{Ar}$ annihilation useful to DUNE. We do this by attempting to make some of the same assumptions (roughly) as GENIE, and vice-versa. For instance, we can and do generate events by simulating the annihilation position sourced from a Woods-Saxon distribution within our generator (alongside the more modern version as developed 
in Sec. II); similarly, with little work, we have perturbed the default settings of the GENIE $n\rightarrow\bar{n}$ generator module to utilize a noninteracting local Fermi gas nuclear model along with a full intranuclear cascade. The inclusion of an $\bar{n}$-potential within GENIE has not yet been investigated; implementation of the modern annihilation position probability distribution (Sec.~\ref{sec:argonlifetime}) is currently underway. While none of these comparisons across generators are ever to be \textit{exact}, their approximately equivalent formalism can serve to inform us as to the stability of quantities which characterize the possible final state topologies of a true $n\rightarrow\bar{n}$ signal event with respect to their associated backgrounds. This stability across models, and their interplay with model detector reconstruction, will be studied in detail in future work with DUNE collaborators.

\begin{figure}[ht!]
    \centering
    \includegraphics[width=1.0\columnwidth]{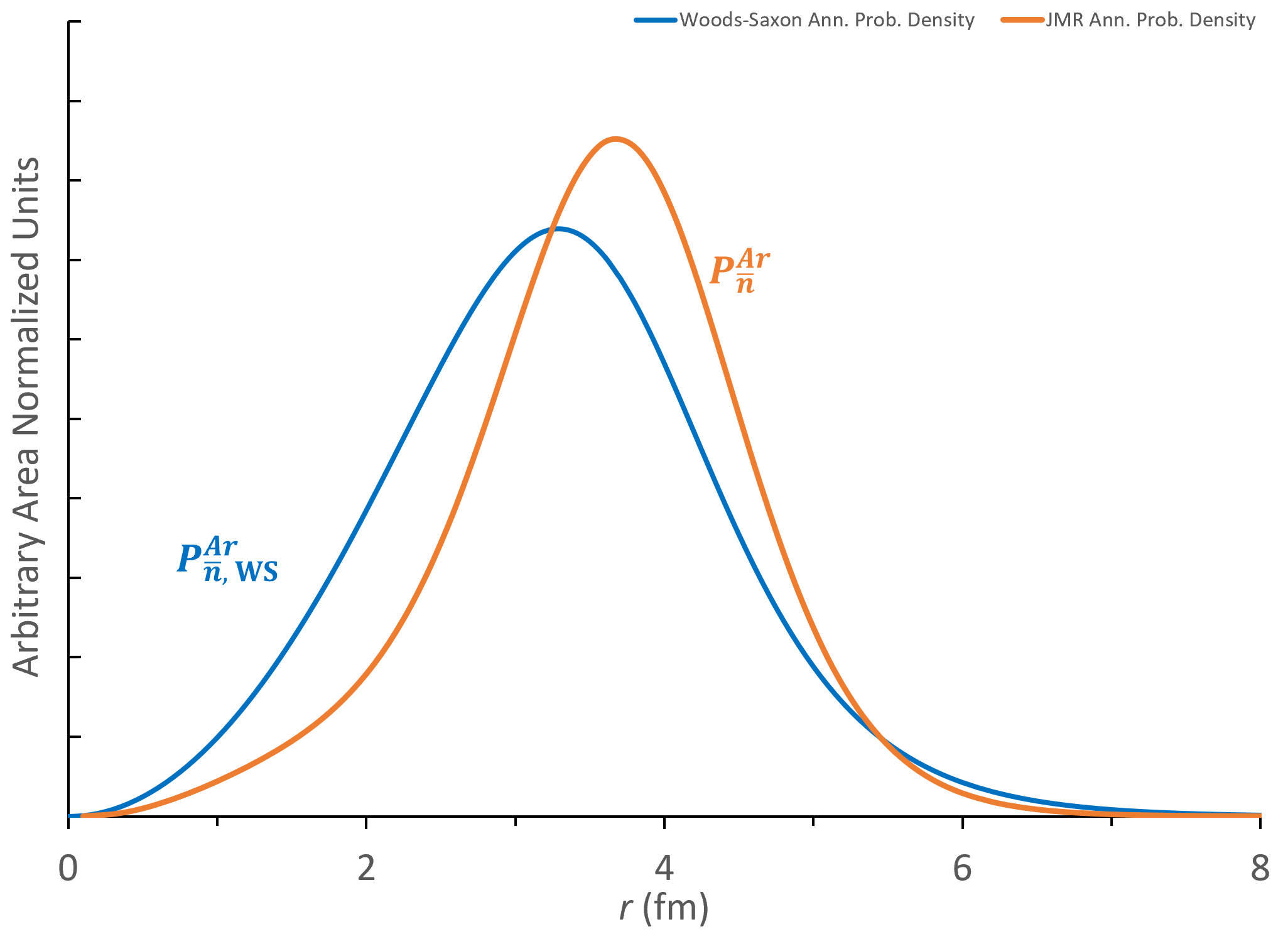}
    \caption{Two plots are shown for various generator assumptions. In blue, we see the naive intranuclear radial position of annihilation probability distribution generated by a Woods-Saxon (abbreviated ``WS"), as presented in GENIE. In orange, we see the modern, quantum-mechanically derived, shell-averaged, true intranuclear radial position of annihilation probability distribution as developed in Sec.~\ref{sec:intranuclear}, present in our generator. Probability distributions are normalized to the same arbitrary integral for a direct comparison, and the scale is arbitrary.}
    \label{fig:AnnDist}
\end{figure}

Two of the probability distributions of intranuclear radial position upon annihilation for these generators are shown in Fig.~\ref{fig:AnnDist} in orange and blue, and are surprisingly similar even with quite different physical assumptions. The quantum mechanical, shell-by-shell distributions discussed in Fig.~\ref{fig:Ar1} are all taken in a weighted average to create the final orange curve, from which our generator can source its initial annihilation positions in a binned fashion; this position is thrown, a nearest neighbor nucleon found within a given zone (and then "moved" to the annihilation position), a Fermi momentum computed for each of the pair given their initial positions, and then a total phase space calculated. All GENIEv3.0.6 events source their annihilation positions from a smooth Woods-Saxon (nuclear density) distribution \citep{Andreopoulos:2015wxa} incredibly similar to our own continuous parameterization for the nuclear density, from which we derive our local zone densities:
\begin{equation}
    \label{eq:smoothwoodssaxon}
    \rho_{WS}^{Ar} = (1+\exp({\frac{r-3.6894}{0.5227}}))^{-1};
\end{equation}
GENIE also similarly throws momenta from (non)local single nucleon momentum distributions. When this curve is multiplied by $r^2$, one generates the blue annihilation probability distribution. These curves effectively demonstrate how even the most simple of assumptions, some only quasi-classical, can lead to quite good approximations; however, we note the preponderance of events \textit{even further} toward or beyond the surface of the nucleus using a quantum-mechanical formalism. The increased likelihood of such surface annihilations, along with their associated correlation with lower momenta and higher final state meson multiplicity, will be shown in the coming figures to be an arguably critical part in the proper evaluation of future experimental efficiencies and possible lower limits on mean intranuclear $n\rightarrow\bar{n}$ transformation time; the interplay of these quantities and any changes in the final state $\pi$-star topology observable in the DUNE detectors has not yet been completely investigated. For similar plots and discussions, see \citep{Golubeva:1992tr,Golubeva:1996ij,Golubeva:1997mc,Golubeva:1997fs,Golubeva:2018mrz}.

\begin{figure}[ht!]
    \centering
    \includegraphics[width=1.0\columnwidth]{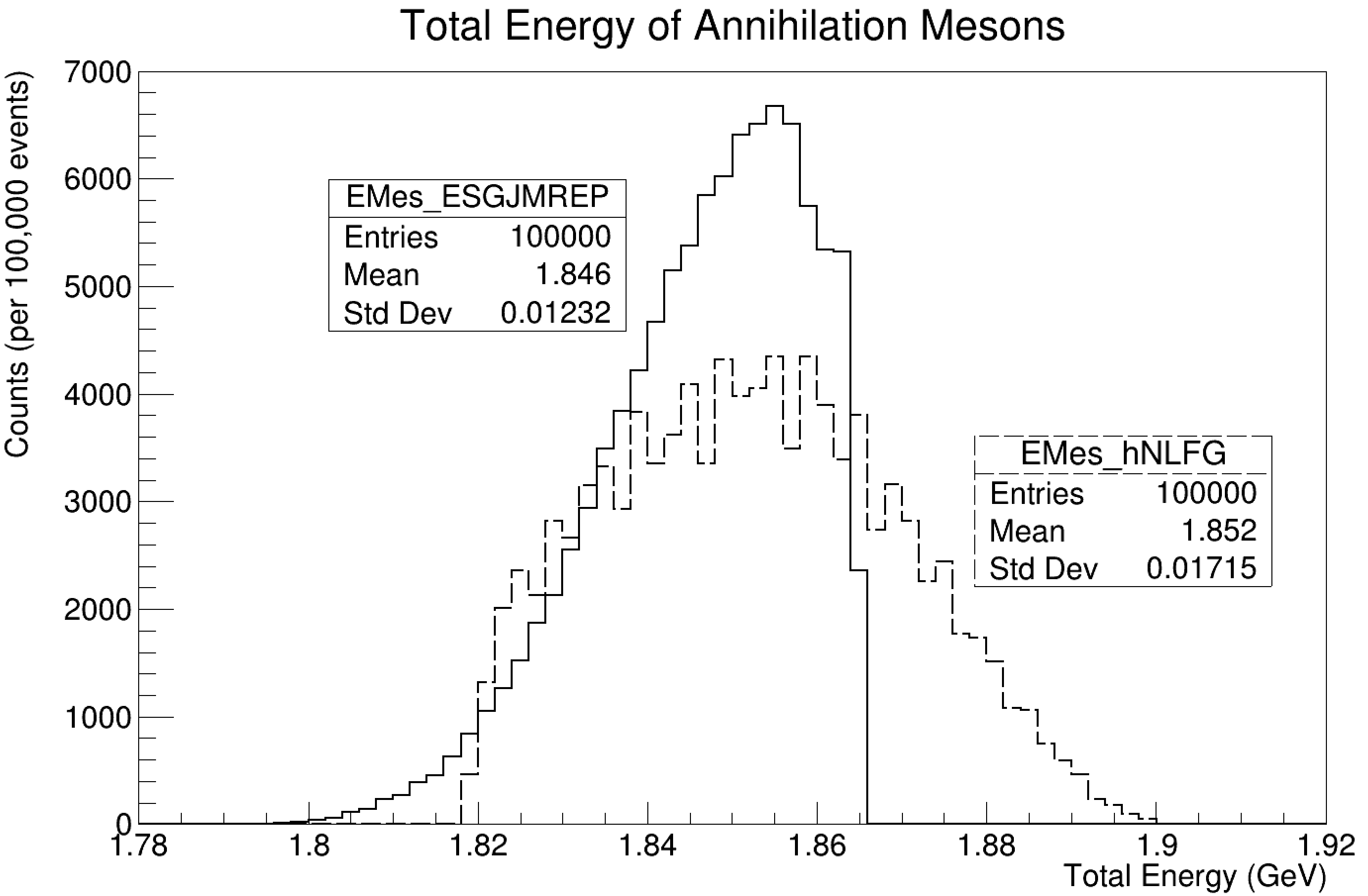}
    \caption{The distributions of total initial annihilation meson energy are shown for this work (solid line) and GENIEv3.0.6 (dashed line) using local Fermi gas models. Via conservation, each of these is equivalent to the distributions of the annihilating $\bar{n}N$ pair.}
    \label{fig:AnnMesTotEDists}
\end{figure}

Some of the differences between this work and GENIEv3.0.6's generator pertain to the initial dynamics of the intranuclear annihilation. An example of this can be seen in Fig.~\ref{fig:AnnMesTotEDists}, showing the initial annihilation mesons' total energy for this work (using the orange curve in Fig.~\ref{fig:AnnDist}) and GENIE (using the blue curve in Fig.~\ref{fig:AnnDist}), each using a version of a local Fermi gas nuclear model. Like the extranuclear $\bar{n}$ annihilation described on $\isotope[12]{C}$ above, energy balance in the annihilation point is given as $E_{ann}+E^{*} = E^{'}_{\bar{n}} + E^{'}_{N}$; taking into account that we have $E^{'}_{\bar{n}}=E^{'}_{n}$, and that
\begin{equation}
    E^{*} = T^{i}_{Fn} - T^{i}_{Fn} + T^{i}_{FN} - T^{i}_{N}
\end{equation}
we have the total energy
\begin{equation}
    E_{ann} + E^{*} = m_{\bar{n}} + m_{N} - 2\epsilon = 1.866\,\mathrm{GeV}\,.
\end{equation}

\begin{figure}[ht!]
    \centering
    \includegraphics[width=1.0\columnwidth]{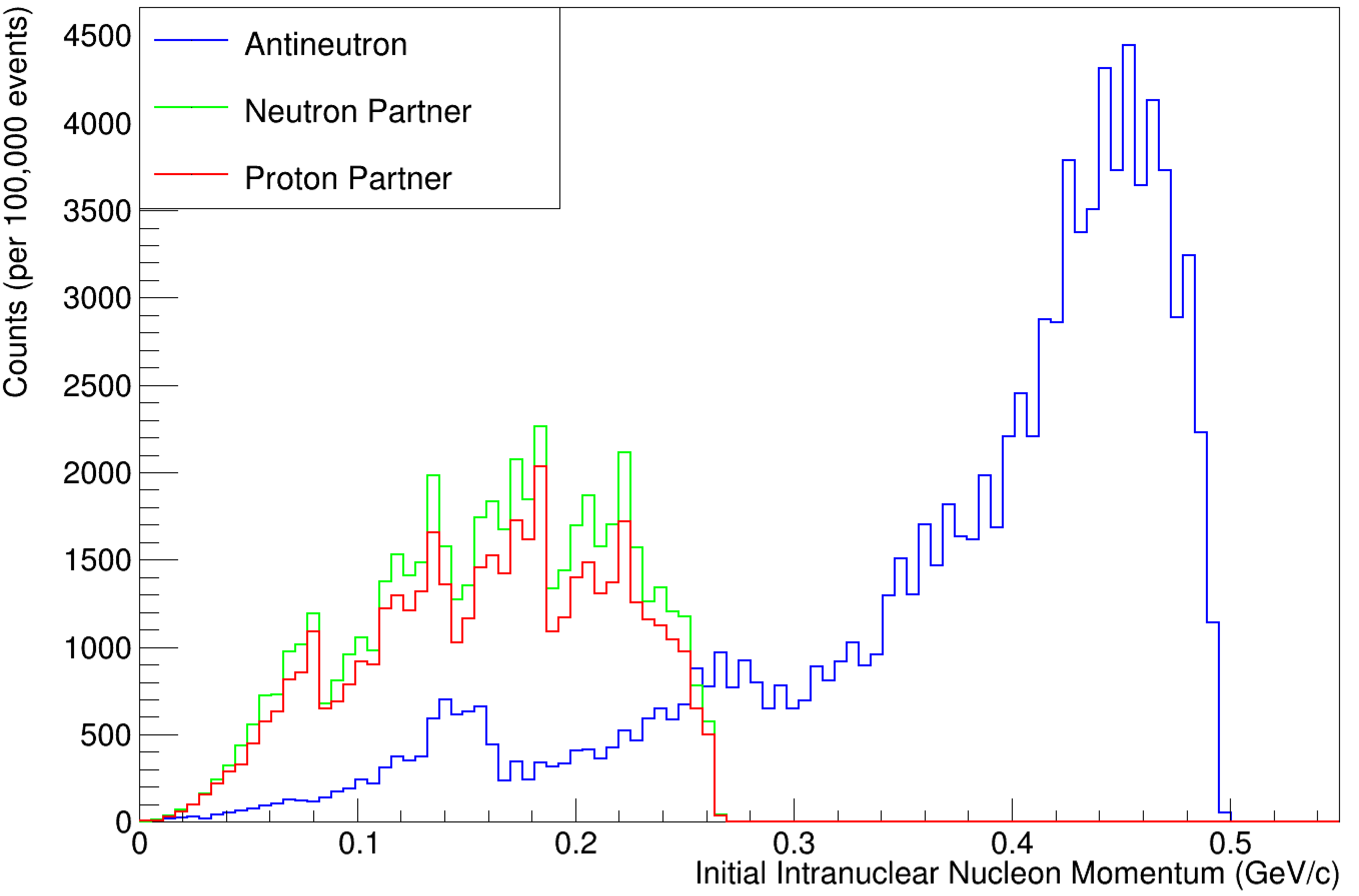}
    \includegraphics[width=1.0\columnwidth]{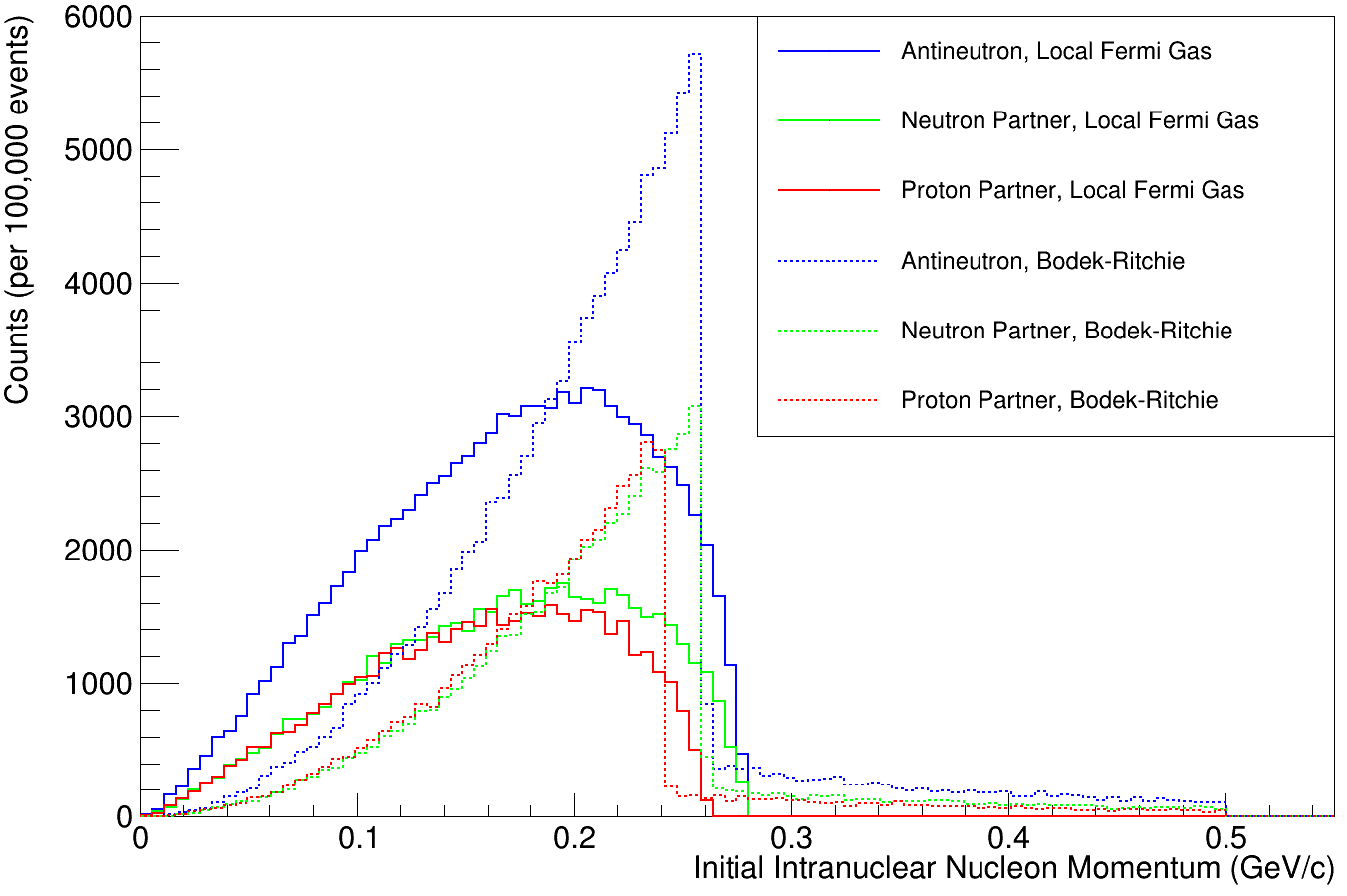}
    \caption{Top: This work, showing the initial (anti)nucleon momentum distributions, using a \textit{zoned} local Fermi gas model with an additional $\bar{n}$ potential. Bottom: the same for the GENIEv3.0.6, showing a local Fermi gas model and the default nonlocal Bodek-Ritchie model.}
    \label{fig:nbarnpP}
\end{figure}

It can be seen in Fig.~\ref{fig:AnnMesTotEDists} that our distribution of energies available to an annihilation is always less than $1.866$\,GeV. GENIE's distribution (which assumes a similar binding energy per nucleon as our model) can be explained simply by considering the minimum/maximum potential magnitudes of annihilation pair momentum (corresponding to an anti/co-parallel intranuclear collision) while assuming an approximately constant intranuclear defect nucleon mass of $\sim 910$\,MeV/$c^2$. The sharp rise of the GENIE distribution around $1.82$\,GeV can be seen to correspond to the addition of momentum distribution shapes around zero momentum (see Fig.~\ref{fig:nbarnpP}).

One can see the different initial single nucleon momentum assumptions in Figs.~\ref{fig:nbarnpP}. The GENIE nonlocal Bodek-Ritchie or local Fermi gas nuclear models mentioned here serve effectively as a set of initial conditions which enable certain nucleon momentum and radial position correlations (or lack thereof). The nonlocal Bodek-Ritchie has been considered the default operating model for $n\rightarrow\bar{n}$ simulations; however, for most of the rest of this article, we will compare local Fermi gas models to each other for a succinct simplicity. Note the different characteristic ranges of momenta; in general, the shapes and ranges of each nucleon local Fermi gas model (solid lines, top and bottom figure) are incredibly similar, while the nonlocal Bodek-Ritchie is quite unique (dashed lines), especially with its phenomenological, short-range ``correlation" tail. For GENIE, we see that the \textit{shapes} of all distributions are identical: $p_f(\bar{n})=p_f(n)\approx p_f(p)$; this is not the case for our model, where $p_f(\bar{n})\neq p_f(n)\approx p_f(p)$ due to the $\bar{n}$-potential. \footnote{For simplicity, throughout this section we have labeled certain plots with ``Golubeva-Richard-Paryev" (for original work done with the modern shell model-derived annihilation position probability distribution and modification of the nuclear medium due to $\bar{n}$ interactions), ``Golubeva-WS-Paryev" (for original work done with a Woods-Saxon-derived annihilation position probability distribution and nuclear medium modification), and then the various GENIEv3.0.6 model identifiers \citep{Andreopoulos:2015wxa}.}

\begin{figure}[ht!]
    \centering
    \includegraphics[width=0.8\columnwidth]{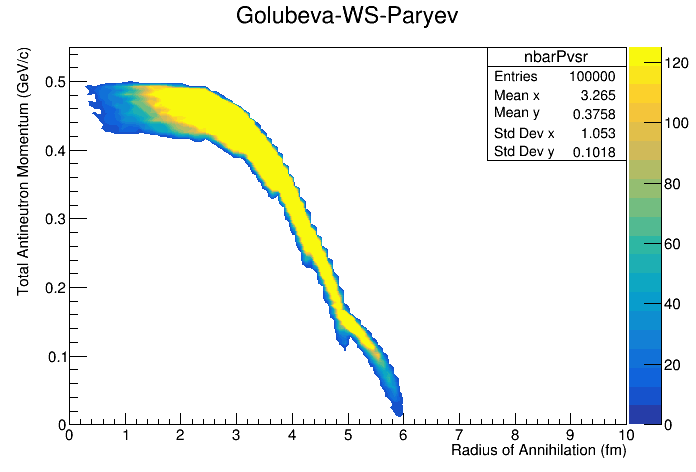}
    \includegraphics[width=0.8\columnwidth]{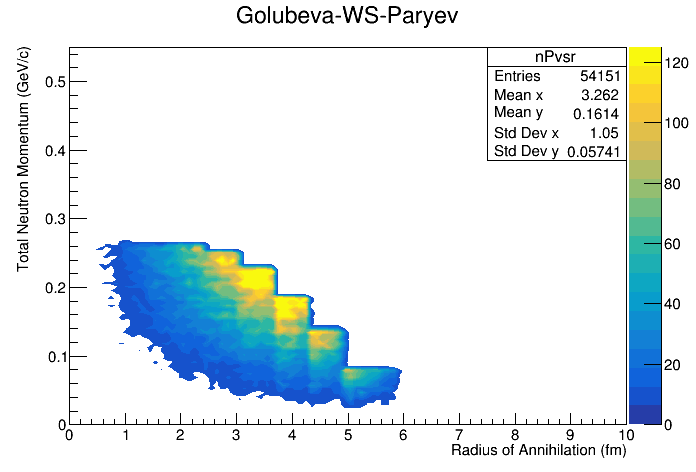}
    \includegraphics[width=0.8\columnwidth]{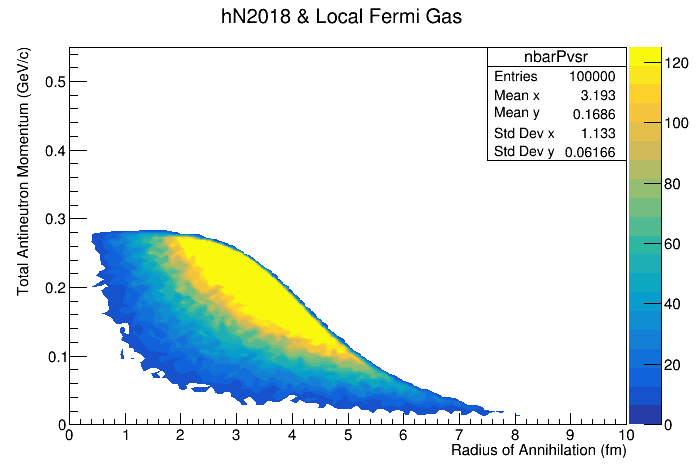}
    \includegraphics[width=0.8\columnwidth]{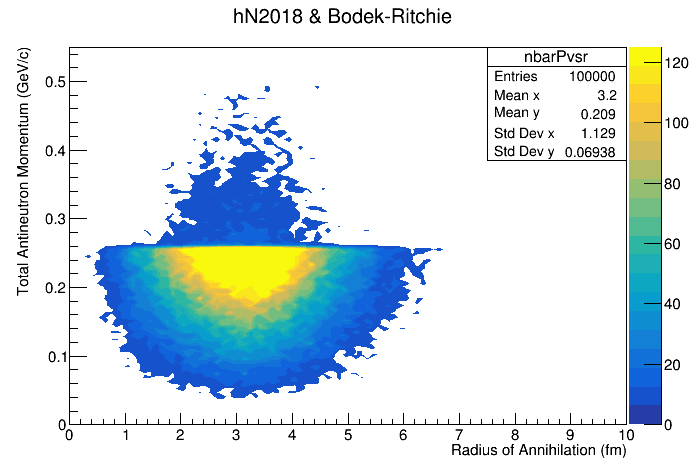}
    \caption{Two dimensional $\bar{n}$ and $n$ momentum-radius correlation plots for this work (top two plots, using a \textit{zoned} local Fermi gas and \textit{zoned} Woods-Saxon annihilation position distribution) alongside GENIEv3.0.6's \textit{local} Fermi gas and \textit{nonlocal} Bodek-Ritchie single (anti)nucleon momentum nuclear models (bottom two plots, also with a smooth Woods-Saxon initial $\bar{n}$ annihilation position distribution).}
    \label{fig:nbarPvsRad}
\end{figure}

The most important aspect of correlated behavior which has been previously unaccounted for in GENIE-affiliated work on $n\rightarrow\bar{n}$ is that of initial (anti)nucleon momentum and radius. In Figs. \ref{fig:nbarPvsRad}, we show comparisons between our and GENIE's local Fermi gas nuclear models, which by definition preserves these correlations,  alongside the inherently \textit{nonlocal} Bodek-Ritchie nuclear model. All figures assume a (zoned or smooth) Woods-Saxon-like annihilation position probability distribution for easier direct comparisons; all outputs from our simulations using the modern annihilation probability distribution from Sec.~\ref{sec:intranuclear} appear somewhat similar when graphed in these coordinates, and so are not shown here to conserve space. The $\bar{n}$-potential is apparent in the top plot, which appears smoothed and lacking of any zoned discontinuities due to the strength of the interaction (as seen in the middle plot, showing correlations for $\bar{n}$ annihilation partners). All \textit{local} models correctly predict a falling-off of nucleon momentum at higher radii, a key consideration for event reconstruction and background rejection. This behavior is not present within GENIE when using the default \textit{nonlocal} Bodek-Ritchie nuclear model, shown at the bottom; the asymmetry present in this plot is due to the Fermi momentum cutoff, above which only phenomenologically 'short-range correlated' $\bar{n}$'s populate.

\begin{figure}[ht!]
    \centering
    \includegraphics[width=1.0\columnwidth]{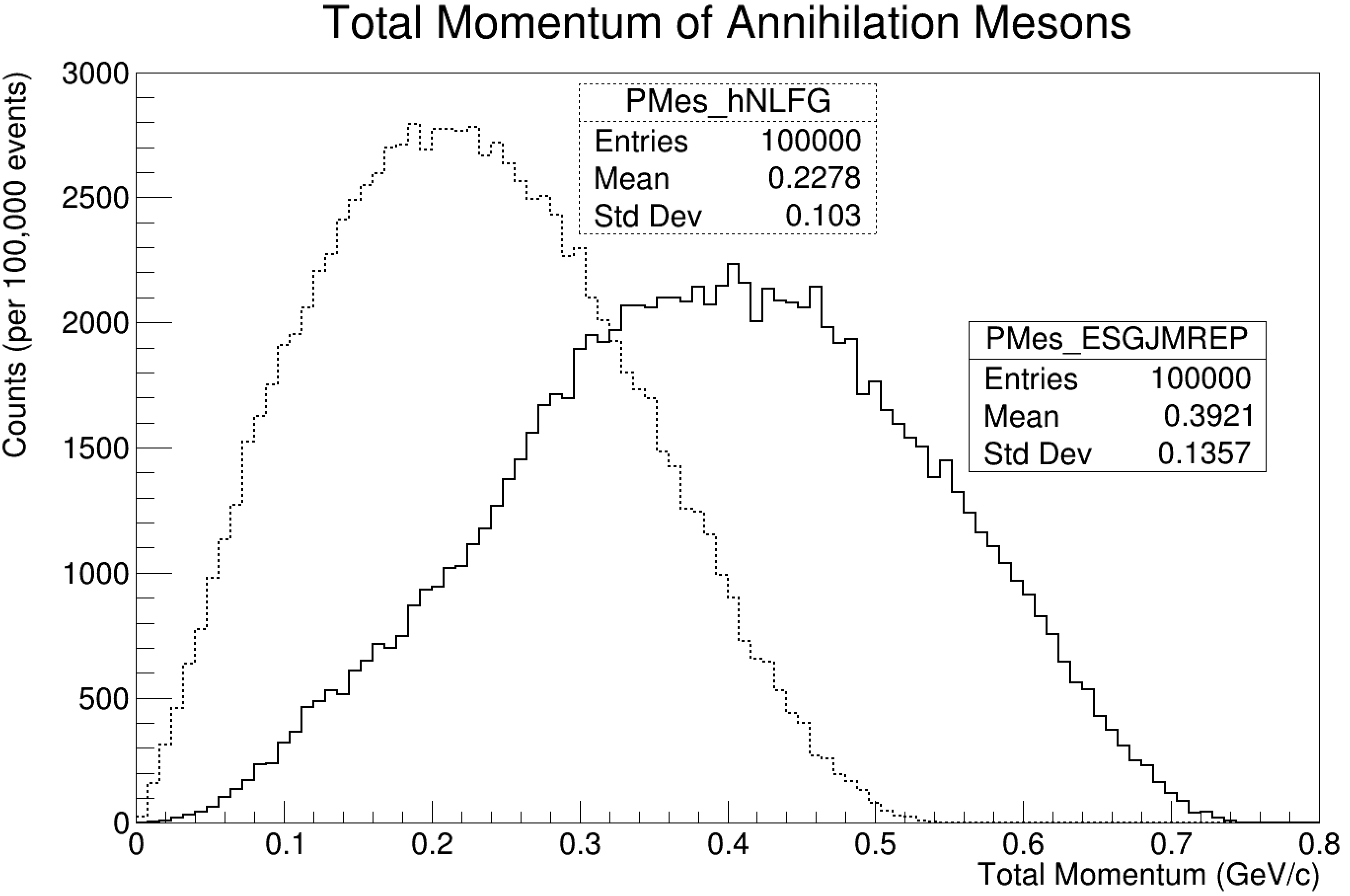}
    \caption{The distributions of total initial annihilation meson momentum are shown for this work (solid line) and GENIEv3.0.6 (dashed line) using local Fermi gas models.}
    \label{fig:AnnMesTotPDists}
\end{figure}

The initial annihilation meson total momentum distribution is seen for our and GENIE models in Fig.~\ref{fig:AnnMesTotPDists}; these distributions are equivalent to the initial two-body annihilation pair momentum distributions by conservation. Each histogram shares a Gaussian-like shape due to the randomized momentum selection from underlying distributions, though our output shows much higher available momentum due to the interaction of the $\bar{n}$ with the modified nuclear medium.

\begin{figure}[h!]
    \centering
    \includegraphics[width=1.0\columnwidth]{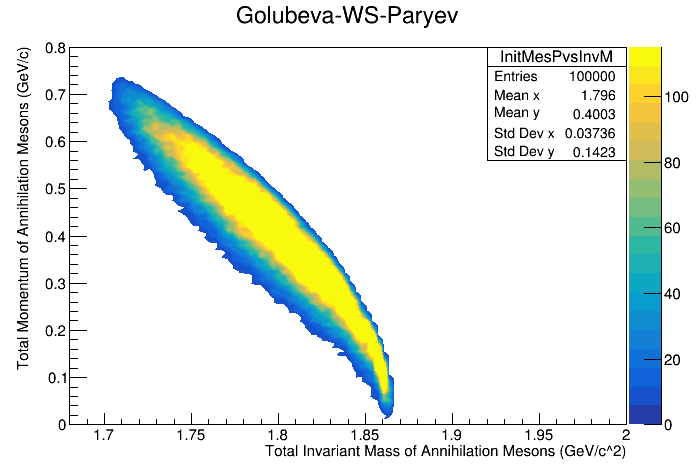}
    \includegraphics[width=1.0\columnwidth]{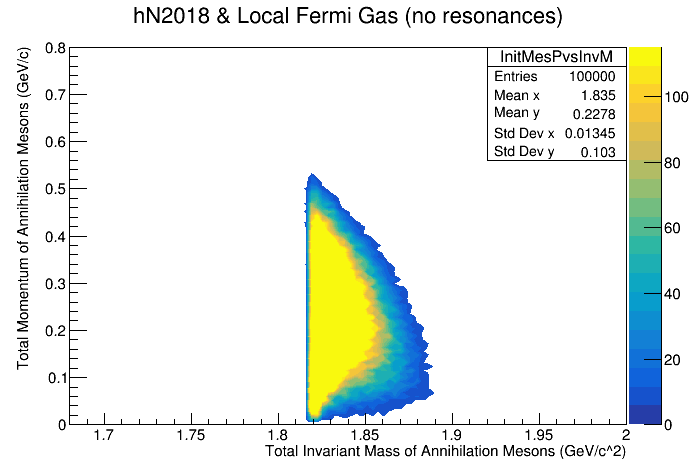}
    \includegraphics[width=1.0\columnwidth]{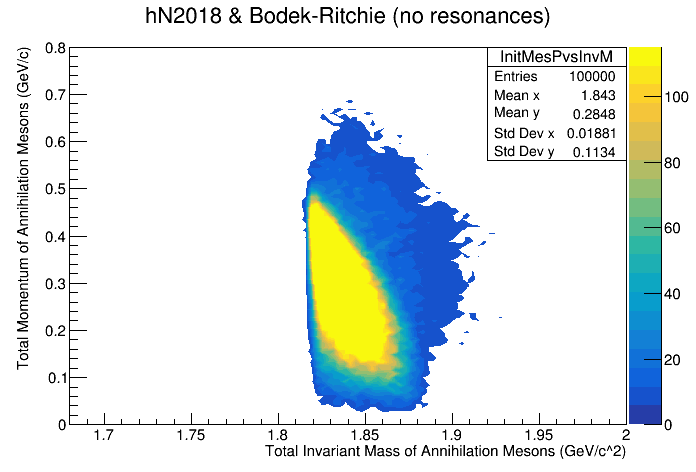}
    \caption{The initial mesonic parameter space (total momentum versus invariant mass) is compared for multiple generators; top, this work; bottom, GENIEv3.0.6. The ``no resonances" phrase refers to a GENIE Mother particle status code cut which removes virtual contributions to invariant mass.}
    \label{fig:InitMesonTotPvInvMass}
\end{figure}

All of this leads us to consider the available initial (and final) mesonic parameter space, again \`a la Fig. 2 of \citep{Abe15}. As seen in Figs.~\ref{fig:InitMesonTotPvInvMass}, this serves as an initial condition of the annihilation-generated mesons before intranuclear transport; thus, for GENIE, hA/hN2018 models (see Secs. 2.5.4 and 2.5.5 of the GENIE v3.0 manual \citep{Andreopoulos:2015wxa} for full discussions) are at this stage equivalent. Due to the \textit{non-dynamical} off-shell masses of annihilation pairs, GENIE predicts higher invariant masses, while our model shows them decrease due to off-shell mass defects in correlation with radial position. Overall, the space is quite differently filled for each model, though considering this is \textit{before} the intranuclear cascade, one cannot necessarily predict much about the final state.

\begin{figure}[h!]
    \centering
    \includegraphics[width=1.0\columnwidth]{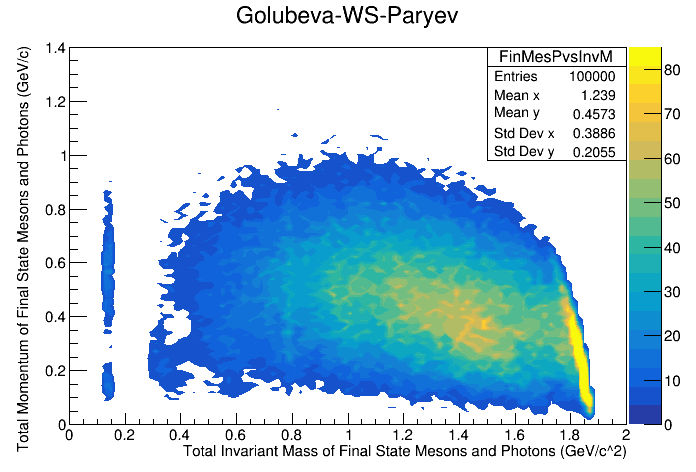}
    \includegraphics[width=1.0\columnwidth]{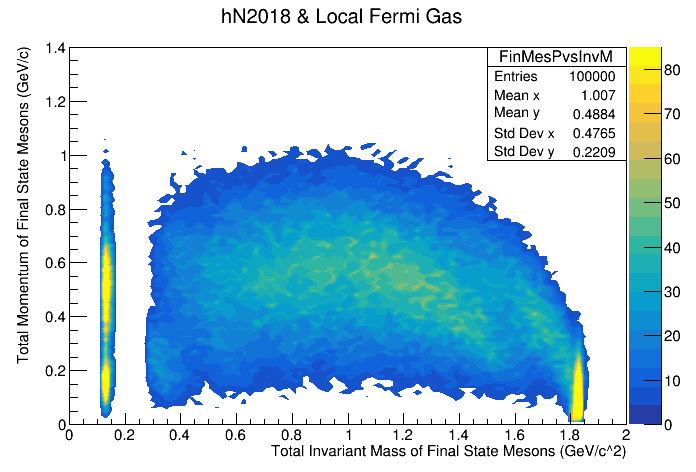}
    \includegraphics[width=1.0\columnwidth]{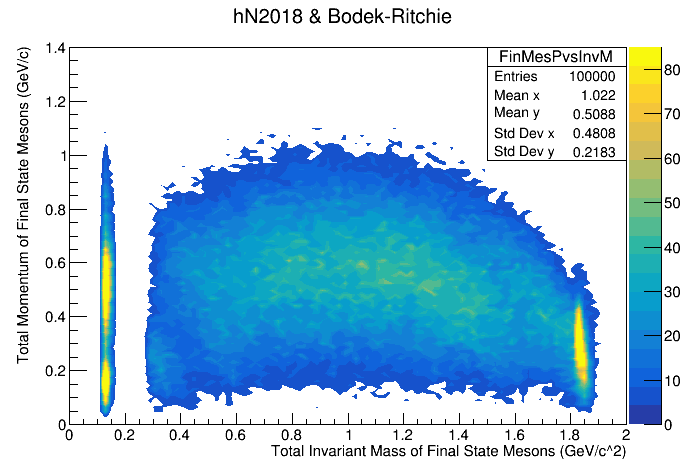}
    \caption{The final state mesonic/pionic parameter space (total momentum versus invariant mass) after intranuclear transport is compared for multiple generators; top, this work; middle and bottom, GENIEv3.0.6 local Fermi gas and nonlocal Bodek-Ritchie, respectivley. Differences in these may lead to different detector signal efficiencies.}
    \label{fig:FinMesonTotPvInvMass}
\end{figure}

The follow-up to these figures can be studied in the comparison of Figs.~\ref{fig:FinMesonTotPvInvMass}, where our and GENIE generated events proceed through a full intranuclear cascade (hN2018 only is shown here). Note that our model includes photons in the final state from high-mass resonance decays, while GENIE does not. The disconnected regions toward the left of the plots are signs of single $\pi$ emission after at least one or more meson absorptions. While \textit{overall} the distribution of events is rather consistent, critically, the high density of events with large invariant mass and low momentum (bottom right of plots) among these local Fermi gas models shows the importance of modeling correlations between position and momentum as they imply a comparatively large number of escaping (and possibly visible) $\pi$'s in the final state. It is with these areas that one may hope to find a significant rejection of background events, possibly allowing for an actual observation of an $n\rightarrow\bar{n}$ event. A full characterization of these effects across many nuclear model configurations within the DUNE detectors is forthcoming within the DUNE Nucleon Decay Group.

\begin{figure}[ht!]
    \centering
    \includegraphics[width=1.0\columnwidth]{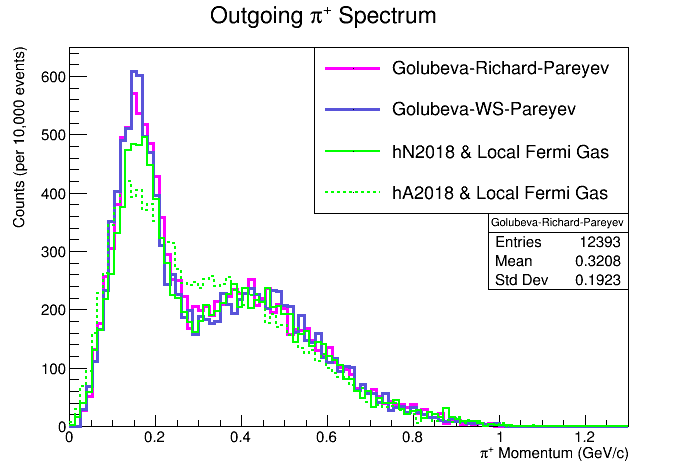}
    \caption{The outgoing $\pi^+$ momentum spectrum is shown for several local Fermi gas models.}
    \label{fig:PiPlusSpectrum}
\end{figure}

\begin{figure}[ht!]
    \centering
    \includegraphics[width=1.0\columnwidth]{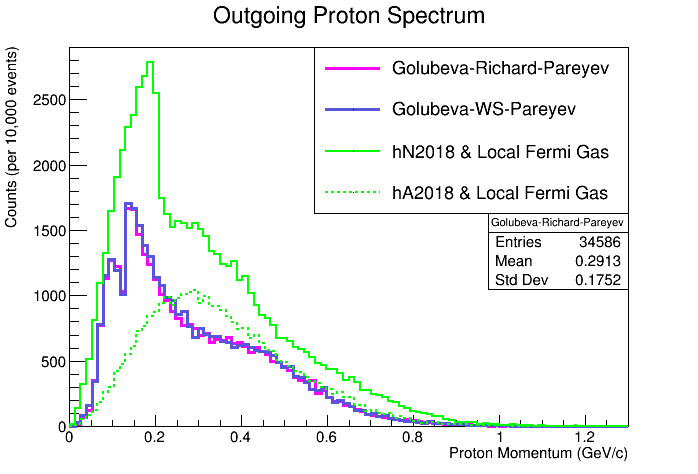}
    \caption{The outgoing $p$ momentum spectrum is shown for several local Fermi gas models.}
    \label{fig:ProtonSpectrum}
\end{figure}

We end the comparison of these generators with Tab.~\ref{tab:final-multiplicities-argon} and Figs.~\ref{fig:PiPlusSpectrum} and \ref{fig:ProtonSpectrum}, which show many similarities and some differences across them. Multiplicities in Tab.~\ref{tab:final-multiplicities-argon} are seen to be most different between our model and GENIE in the realm of outgoing nucleons (resulting from nucleon knock-out or evaporative de-excitation); this should not be surprising, as GENIE does not currently contain a public version of an evaporation model within either its hN2018 (full intranuclear cascade) or hA2018 (single effective interaction) models. Exciting work by GENIE developers in this regime is expected to be completed soon, and we look forward to being able to compare our results. Small differences can also be seen in the $\pi^{0,\pm}$ multiplicities, which are partially due to the fact that we predict more $n\rightarrow\bar{n}$ events toward or beyond the surface of the nucleus (see the blue and orange radial annihilation position probability distributions of Fig.~\ref{fig:AnnDist}), but there are also nontrivial dependencies given the larger number of possible branching channels we simulate \citep{Golubeva:2018mrz} compared to GENIE \citep{Andreopoulos:2015wxa}.

\onecolumngrid

\begin{center}
\begin{table}[ht!]
\caption{Final state average stable particle multiplicities for several 10,000 event samples across multiple $\bar{n}Ar$ annihilation MCs. Also included is the initial total annihilation energy. See \citep{Golubeva:2018mrz} for a full description of the zoned local Fermi gas and the intranuclear cascade used in this work. See the GENIE v3.0 manual \citep{Andreopoulos:2015wxa} for discussions of nuclear models and intranuclear cascades.}
\label{tab:final-multiplicities-argon}
\begin{ruledtabular}
    \begin{tabular}{l|ccccccc}
     & $M(\pi)$ & $M(\pi^+)$ & $M(\pi^-)$ & $M(\pi^0)$ & $M(p)$ & $M(n)$ & $E_{o}^{tot}$\,(MeV)\\ \hline
    $\bar{n} {}^{39}_{18} {\rm Ar}$ Golubeva-Richard-Paryev (\textit{Zoned} Local Fermi Gas, INC) & $3.813$ & $1.239$ & $1.008$ & $1.566$ & $3.459$ & $4.823$ & $1.846$\\
     $\bar{n} {}^{39}_{18}{\rm Ar}$ Golubeva-Woods-Saxon-Paryev (\textit{Zoned} Local Fermi Gas, INC) & $3.781$ & $1.22$ & $0.998$ & $1.563$ & $3.63$ & $4.896$ & $1.845$\\
    $\bar{n} {}^{39}_{18} {\rm Ar}$ GENIEv3.0.6 (\textit{Default} Bodek-Ritchie, hA2018) & $3.610$ & $1.183$ & $0.991$ & $1.436$ & $3.021$ & $3.151$ & $1.925$\\
     $\bar{n} {}^{39}_{18} {\rm Ar}$ GENIEv3.0.6 (\textit{Default} Bodek-Ritchie, hN2018) & $3.280$ & $1.159$ & $0.968$ & $1.153$ & $6.192$ & $6.654$ & $1.922$\\
    $\bar{n} {}^{39}_{18} {\rm Ar}$ GENIEv3.0.6 (Local Fermi Gas, hA2018) & $3.594$ & $1.173$ & $0.9776$ & $1.444$ & $3.045$ & $3.174$ & $1.908$\\
    $\bar{n} {}^{39}_{18} {\rm Ar}$ GENIEv3.0.6 (Local Fermi Gas, hN2018) & $3.24$ & $1.155$ & $0.956$ & $1.129$ & $6.269$ & $6.718$ & $1.905$\\
    \end{tabular}
\end{ruledtabular}
\end{table}
\end{center}

\twocolumngrid

To give a more complete context to Table~\ref{tab:final-multiplicities-argon}, we plot several (absolute magnitude) final state momentum spectra for $\pi^+$ and $p$ species, two key constituents in the eventual experimental search for $n\rightarrow\bar{n}$ in DUNE; note that the $\pi^{0,-}$ distributions are quite similar. In Fig.~\ref{fig:PiPlusSpectrum}, we see that our models agree quite well with the full intranuclear cascade simulation from GENIE (using hN2018) in both multiplicity and shape; there is some lack of structure around the $\Delta$-resonance within the hA2018 simulation (recall this models the cascade as a \textit{single effective interaction} using tabulated reaction rates), a sign of the competition between cross-sections (or rates) of processes such as $\Delta$ decay and pion absorption.

In Fig.~\ref{fig:ProtonSpectrum}, we see the outgoing proton spectra. Here, our model and GENIE differ greatly (and GENIE even among itself) across lower momenta. Though there is a full intranuclear cascade model within GENIE (hN2018), it can be directly seen that it does not yet include any nucleon evaporation \textit{currently}. In some respect, these differences should be expected due to the novel nature of the phenomena we are modeling (transporting $\sim 4\mbox{-}5$ mesons is no easy business), and the fact that our generator was comparatively purpose-built to reproduce antinucleon annihilation data. We will note, however, that if one takes a more experimental viewpoint, these are not actually so disparate; indeed, if we consider an \textit{approximate, conservative, minimum} proton kinetic energy detectability threshold in liquid \Arl ~to be $\gtrsim 100$\,MeV (i.e., $\gtrsim 450$\,MeV$/c$) \citep{MicroBooNE1025}, we see that above this value much of the shape and magnitude of \textit{all} distributions are quite similar. Thus, in some respect, we expect these models to appear rather degenerate for protons when taking detector response into account. 

\section{\label{sec:conclusion}Conclusions and Outlook}

In this work, we have endeavored to give an update to the community on recent developments in the modeling of $\bar{n}A$ annihilation events in service of future BSM $n\rightarrow\bar{n}$ searches. Critical among the findings of this paper is the calculation of a new intranuclear suppression factor for \Arl, $T_R^{Ar}\sim 5.6\times 10^{22}\,\mathrm{s}^{-1}$, along with a new and associated calculation of the \Arl~intranuclear radial annihilation probability distribution. Also, efforts have been made to implement this and other important $\bar{n}N$ annihilation dynamics into an independently developed, antinucleon-data-driven MC generator to service both the ESS NNBar Collaboration and DUNE. Samples of 100,000 events for both communities are now being prepared and will be made available upon reasonable request to the authors. Comparisons and discussions of differences and similarities have been made to data where available, previous MC results, and other publicly available event generators such as GENIE.

Within this work, a kind of forward path has been illuminated for the (intranuclear) BSM (di)nucleon decay community, showing the importance of \textit{some} initial physical correlations in the modeling of BSM signals, most importantly that of the event constituent's momenta and intranuclear position. However, the effects of final state interactions cannot be understated. It is with these findings, and our associated event generator, that we hope to empower future experiments to better understand probable signal topologies for rare decays. Collaboration is ongoing with DUNE community members within the Nucleon Decay Working Group to study the implications of this modeling work on possible efficiencies, atmospheric neutrino background rejection rates, and lower limits on the $n\rightarrow\bar{n}$ mean transformation time inside simulated DUNE detector geometries.

Another future step elucidated by this work is the critical nature of current and future collaborations' endeavors to holistically evaluate and compare various $\bar{n}A$ interactions using common formalisms for the calculation of (anti)nucleon wave functions, radial annihilation probability distributions, and intranuclear suppression factors for all pertinent nuclei. This is a rather monumental task, but one which should be completed in the same way as for other rare decay searches, such as within the $0\nu\beta\beta$ community.

\begin{acknowledgments}

JMR would like to thank Michael Bender and Karim Bennaceur for useful  advice about shell-model wave functions, Joseph Cugnon for discussions, and Dario Auterio and his collaborators for interesting discussions about the DUNE experiment. JLB's work is supported by the U.S. Department of Energy, Office of Science, Office of Workforce Development for Teachers and Scientists, Office of Science Graduate Student Research (SCGSR) program. The SCGSR program is administered by the Oak Ridge Institute for Science and Education for the DOE under contract number DE‐SC0014664. JLB is grateful to the US DOE SCGSR Program and Fermilab for their generous support of this and other work. ESG would like to thank INR, Moscow, for the support of her travel.
We all wish to thank the organizers of the INT Neutron Oscillations Workshop in Seattle during November 2017, the ILL-LPSC PPNS Workshop at Grenoble in May 2018, and the ESS Workshop at NORDITA in December 2018 for providing a stimulating environment where these topics have been thoroughly discussed. We also wish to thank Yuri Kamyshkov for his constant stimulus and interests in this work, helping to drive us all along.
\end{acknowledgments}



\end{document}